\documentclass{iopjournal}

\usepackage{xcolor}
\usepackage{graphicx}
\usepackage{dcolumn}
\usepackage{bm}
\usepackage[utf8]{inputenc}
\usepackage[T1]{fontenc}
\usepackage{mathptmx}
\usepackage{amssymb}
\usepackage[version=4]{mhchem} 
\usepackage{caption}
\usepackage{subcaption}
\usepackage{natbib}
\usepackage{float}
\usepackage{url}
\usepackage{multirow}
\usepackage{array}
\usepackage{wrapfig}
\usepackage{tabularx}
\usepackage{color}
\usepackage{siunitx}
\usepackage{hyperref}

\usepackage{verbatim}
\usepackage{comment}
\usepackage{gensymb}

\begin{document}

\articletype{Paper}

\title{Plasma morphology under nanosecond repetitively pulsed discharges with air cross-flow: A simulation study}

\author{Gihun Shim$^1$\orcid{0000-0003-2129-5197}, Praise Noah Johnson$^1$\orcid{0000-0002-2212-7147}, Taaresh Sanjeev Taneja$^1$\orcid{0000-0002-0233-2039}, and Suo Yang$^{1,*}$\orcid{0000-0002-9924-2645}}

\affil{$^1$ Department of Mechanical Engineering, University of Minnesota -- Twin Cities, Minneapolis, MN 55455, USA}

\affil{$^*$ Author to whom correspondence should be addressed.}

\email{suo-yang@umn.edu}


\begin{abstract}
Nanosecond repetitively pulsed discharges in air cross-flows exhibit distinct plasma morphologies, which can expand the spatial extent of the bulk plasma regions and extend the residence time of reactive radicals within the flow field. This work develops a high‑fidelity plasma–fluid simulation framework within OpenFOAM, which couples drift–diffusion–convection transport, electrostatic potential, photoionization, mixture-averaged transport, ambipolar diffusion, and a local mean electron energy formulation. Using this framework, we systematically investigate plasma morphologies under air cross-flow velocities of $40$ and $100\text{ m/s}$ at a $50\text{ kHz}$ repetition frequency. The simulations reveal that lower cross-flow velocity promotes plasma retention within previously formed channels, resulting in substantial thermal accumulation. Conversely, at higher cross-flow velocities, plasma is rapidly convected downstream, reducing plasma retention between pulses, and weakening thermal accumulation. The simulations further reveal the physical mechanisms underlying snap-back, in which the subsequent discharge shifts from a previously formed channel toward a shorter, more direct path between the electrodes as the conductivity of the previous channel decreases. When a strong electric field forms between the electrodes, it accelerates electrons and causes intense Joule heating near the anode tip. This increases the local electron energy, enhancing a production rate of electrons and generating a stronger current between the electrodes. As the conductivity of the previous channel decreases, the discharge consequently establishes a new path of least resistance directly between the electrodes.
\end{abstract}

\section{Introduction} 
Nonthermal plasmas have emerged as a cornerstone technology across a diverse spectrum of modern engineering applications, ranging from plasma-assisted combustion (PAC)~\cite{ju2015plasma,ju2016plasma,popov2016kinetics} to the abatement of hazardous pollutants~\cite{zhang2017non,schiavon2017non} and the synthesis of high-value nanomaterials~\cite{kortshagen2016nonthermal,lin2015microplasma}. Unlike thermal plasmas, which operate in local thermodynamic equilibrium (LTE), the nonthermal plasmas are characterized by a substantial energy nonequilibrium~\cite{colombo2008thermodynamic}. In particular, when electrons are accelerated by an external electric field, they gain energy from the field, resulting in an increase in the local mean electron energy, while the heavy ions and neutral species remain near ambient temperatures. This unique thermodynamic state enables the new high-energy chemical pathways, primarily through electron-impact reactions, without the prohibitive energy cost associated with heating the bulk gas~\cite{palma2020review}. Consequently, nonthermal plasmas offer a pathway to "cold" chemistry that is both energy-efficient and highly controllable~\cite{ju2025plasma, chen2017characteristics}. A key feature of nonthermal discharges is the streamer: a transient, filamentary ionization wave whose propagation involves coupled multiscale and multiphysics processes occurring over nanosecond timescales and micrometer spatial scales~\cite{nijdam2020physics,bruggeman2017foundations,huiskamp2020nanosecond}. As the streamer advances, it generates a localized region of intense electric-field at the streamer head, driving rapid electron-impact ionization and subsequent UV-photon emission~\cite{nijdam2020physics}. The photons pre-ionize the gas ahead of the streamer front, facilitating the continued propagation of the discharge. Modeling streamer initiation and propagation is a formidable challenge, as it requires the simultaneous resolution of drift-diffusion, and Poisson's equations for the electric potential. When reacting-flow dynamics are coupled with streamer dynamics, the resulting model becomes more comprehensive by incorporating convection, multi-species transport, and detailed chemical-kinetic mechanisms involving hundreds of chemical reactions.

A particularly intriguing aspect of nonthermal plasmas is the interaction between the plasma and the cross-flow. While standard models can predict the initial breakdown of streamer and its propagation, the physical mechanisms that dictate the eventual length, expansion, and stretching of the plasma filament remain a subject of active debate~\cite{zhao2021elongation}. This evolving plasma morphology is not merely a geometric feature;; it governs the volume of the plasma-activated region and determines the residence time of reactive species within the discharge. Jacobian~\textit{et al.} investigate ignition and flame‑spreading processes in a fuel jet interacting with a supersonic crossflow~\cite{jacobsen2008plasma}. They demonstrate that compact plasma torches can generate localized high‑enthalpy kernels that couple effectively with fuel jets, thereby promoting ignition and stabilizing combustion under scramjet‑relevant conditions. Do~\textit{et al.} experimentally demonstrate that a nanosecond pulsed nonequilibrium plasma discharge placed inside a cavity significantly enhances flame ignition and stabilization in supersonic crossflows~\cite{do2010plasma}. Their results show that the plasma discharge reduces the ignition delay time of hydrogen and ethylene jets, enabling cavity flames to ignite earlier and propagate farther into the freestream than auto‑ignited flames. Ombrello~\textit{et al.} demonstrate that nanosecond‑pulsed, high‑frequency plasma discharges significantly enhance ignition inside a scramjet cavity, even under the extremely short residence times and harsh aerodynamic conditions of supersonic flow~\cite{ombrello2024scramjet}. The plasma produces rapid, localized energy deposition and generates reactive radicals that accelerate fuel–air chemical reactions, enabling ignition in cases where the flow would otherwise remain nonreactive. Shen~\textit{et al.} experimentally investigate nanosecond‑pulsed, high‑frequency discharge ignition in high‑velocity air and methane/air mixtures with flow speeds up to 80 m/s~\cite{shen2025nanosecond}. The experiments show that hot, ionized gas from previous pulses creates a preferential low‑resistance path that allows subsequent discharges to elongate downstream. Therefore, the plasma morphology strongly depends on flow velocity, inter‑pulse time, and electrode gap, and it directly influences ignition probability. Understanding the evolution of plasma morphology under cross-flow is critical for the optimization of plasma-assisted combustion (PAC) and reactor designs.

In this work, we propose a detailed modeling framework for nonequilibrium plasmas coupled with reacting flows. Implemented within the OpenFOAM environment, the solver employs either the local field approximation (LFA) or the local mean energy approximation (LMEA) and is coupled with a reacting-flow framework. In the LFA, the electron energy distribution function (EEDF) is assumed to equilibrate instantaneously with the local electric field, allowing the transport coefficients (such as mobility and diffusion) of charged species to be expressed as functions of the reduced electric field ($E/N$). In contrast, the LMEA determines reaction rate and transport coefficients from the local mean electron energy by solving the electron energy conservation equation. These coefficients are obtained from lookup tables precomputed using BOLSIG+~\cite{hagelaar2016brief}, enabling a more accurate representation of nonlocal electron kinetics. Our integrated plasma-fluid solver incorporates drift–diffusion–convection modeling, the Poisson's equation for electrostatic potential, photoionization models, mixture-averaged transport for species, and a flexible chemical kinetic interface for both plasma and neutral chemistry. This framework enables high-fidelity simulations of streamer initiation and propagation in complex gas mixtures and glow discharges coupled with reacting-flow systems. Specifically, our solver offers the flexibility to simulate multiple pulses by explicitly resolving the pulse and inter-pulse gap phases in complex geometries using unstructured meshes. By leveraging these capabilities, we investigate how plasma morphology and thermal effects evolve with cross-flow velocity in air and analyze the snap-back phenomenon, in which the previously heated and ionized channel loses conductivity, causing the subsequent discharge to shift toward a shorter, more direct path between the electrodes.

This paper is organized as follows. Section~\ref{s_numeric} introduces numerical methods used to model the pulse phase (streamer propagation, glow discharge), and the gap phase for multi‑pulse discharges based on a drift-diffusion-convection model. In Section~\ref{s_elongation}, we investigate effects of cross-flow velocities on plasma morphology, and physical mechanism of snap-back. Section~\ref{s_conclusion} summarizes the main findings of this study.

\section{Numerical methodology} \label{s_numeric}
For drift–diffusion–convection modeling of the avalanche-to-streamer transition, streamer propagation, and afterglow, we employ an in-house OpenFOAM solver~\cite{jasak2009openfoam}, extended to incorporate detailed plasma-chemical kinetics for the accurate representation of plasma-combustion reaction pathways and their coupling with transport phenomena. Our solver, \textit{PlasmaCombustionFoamLM} (or \textit{PCFoamLM}), the low-Mach-number version of our plasma–combustion solver, has been developed to efficiently capture the complex dynamics of reacting flows, preserving the essential coupling among drift, diffusion, and convection. The governing equations for the detailed modeling of the plasma–fluid system are solved to simulate nanosecond repetitive pulsed discharges (NRPDs) under atmospheric conditions. Therefore, two distinct phases, the pulse and gap phases, are implemented in the solver to model multiple pulses and inter-pulse regimes.


\subsection{Governing equations} 
\subsubsection{Pulse phase} 
The plasma–flow system is described by the compressible Navier–Stokes equations coupled with multi‑species transport and energy conservation. Plasma effects are incorporated into the flow solver through conservation equations for charged species and electron energy, along with source terms representing heat release from plasma chemistry and energy deposition by the discharge. A mixture-averaged transport model is used for species conservation in the \textit{reactingDNS} solver~\cite{reactingDNS_OpenFOAM}, an OpenFOAM-based DNS framework for chemically reacting flows. The model accounts for convection and chemical production and is extended in our solver to include drift, diffusion, and photoionization. The sensible‑enthalpy equation includes convective and diffusive heat transfer, viscous dissipation, pressure work, and energy exchange due to chemical reactions and plasma power deposition, coupled with a mean electron energy equation and Poisson’s equation for the self-consistent electrostatic potential. Together, these equations form a detailed framework for simulating reacting plasma flows using a drift–diffusion–convection formulation:
\begin{align}
 \frac{\partial \rho}{\partial t} + \nabla \cdot \left(\rho \vec{u}\right) &= 0, \\
 \frac{\partial \rho \vec{u}}{\partial t} + \nabla \cdot (\rho \vec{u} \vec{u}) &= -\nabla p + \nabla \cdot \tau + \sum_{i=1}^{N_{sp}} Z_i e n_i \vec{E},  \\
 \frac{\partial n_e}{\partial t} + \nabla \cdot \left(\mu_e \vec{E} n_e - D_e \nabla n_e + \vec{u} n_e \right) &= N_A \dot{\omega}_e + S_{ph,e}, \\
 \frac{\partial \rho Y_{i}}{\partial t} + \nabla \cdot \left[ \frac{Z_i}{|Z_i|} \rho \mu_i \vec{E} Y_i - \rho D_i \nabla Y_i + \rho \vec{u} Y_i \right] &= W_{i} \dot{\omega}_{i} + \frac{W_{i} S_{ph,i}}{N_A}, \\ 
 \frac{\partial E_e}{\partial t} + \nabla \cdot \left(\mu_\epsilon \vec{E} E_e - D_\epsilon \nabla E_e + \vec{u} E_e \right) &= -e \vec{\Gamma_e} \cdot \vec{E} - \frac{3}{2} n_e k_B \left(T_e - T_g\right) \frac{2 m_e}{m_g} \nu - \dot{E}_{plas}, \\
 \frac{\partial \rho h_s}{\partial t} + \nabla \cdot (\rho h_s \vec{u} + \vec{q}) + \tau : \nabla \vec{u} &= \frac{Dp}{Dt} + \dot{Q}_{react} + \frac{3}{2} n_e k_B \left(T_e - T_g\right) \frac{2 m_e}{m_g} \nu + \dot{E}_{plas, nB}, \\
 \nabla^2 \phi_e + \frac{e}{\epsilon_0} \sum_{i=1}^{N_{sp}} Z_i n_i &= 0.
\end{align}  
where $\vec{u}$ is the fluid velocity, $t$ the time, $\rho$ the mixture density, $p$ the pressure, $\tau$ the viscous stress tensor, $Z_i$ the charge number of species $i$, $N_{sp}$ the number of species, $e$ the elementary charge, $\vec{E}$ the electric field, and $n_i$ the number density of species $i$. Note that the final term on the right‑hand side of the momentum equation represents the ionic wind, arising from ion drift under the Coulomb force and the resulting momentum transfer to the neutral gas. In the species conservation, $n_e$ denotes the electron number density, $\mu_e$ the electron mobility, and $D_e$ the electron diffusion coefficient. The term $\dot{\omega}_e$ represents the net molar rate of electron production/destruction due to chemical reactions such as ionization, recombination, and attachment, and $S_{ph,e}$ is the photoionization source term for electrons. For heavy species, $Y_i$ is the mass fraction of species $i$, $\mu_i$ is the mobility, $D_{i}$ the effective diffusion coefficient from the binary diffusion coefficients~\cite{zhong2020structure}, $W_{i}$ the molar weight of species $i$, $\dot{\omega}_i$ the net molar rate of ion production/destruction due to chemical reactions, $S_{ph,i}$ the photoionization source term, and $N_A$ the Avogadro’s number. The binary diffusion coefficients for ion-neutral collisions are taken from~\cite{alkandry2013comparison}, with diffusion collision integrals for the ion–neutral pairs obtained from~\cite{hash2007fire}. The local mean electron energy, defined as $E_e = \frac{3}{2} n_e k_B T_e$, also accounts for drift, diffusion, and convection, where $\mu_\epsilon$ and $D_\epsilon$ are the mobility and diffusion coefficient of the mean electron energy, respectively. The right-hand side of the electron energy equation accounts for Joule heating, elastic energy exchange with the background gas, and inelastic energy losses $\dot{E}_{plas}$ due to electron-impact reactions. Here, $\vec{\Gamma}_e (= \mu_e \vec{E} n_e - D_e \nabla n_e)$ the electron flux, $T_e$ the electron temperature, $T_g$ the background gas temperature, $m_e$ and $m_g$ the electron and gas particle masses, and $\nu$ the electron--neutral momentum‑transfer collision frequency, respectively. In the sensible enthalpy, $h_s$ denotes the sensible enthalpy, $\vec{q}$ the molecular heat flux, $\tau : \nabla \vec{u}$ the viscous dissipation, $\dot{Q}_{react}$ the heat source term due to chemical reactions, and $\dot{E}_{\mathrm{plas,nB}}$ denotes the plasma power deposition from non-BOLSIG electron-impact reactions. In Poisson’s equation, $\phi_e$ denotes the electrostatic potential, and $\epsilon_0$ the permittivity of free space.
The chemical heat-release term $\dot{Q}_{react}$, and the deposited energy by plasma discharge $\dot{E}_{plas}$ are given by
\begin{align}
 \dot{Q}_{react} &= -\sum_{i=1}^{N_{sp}} h_i^0 \dot{\omega}_i, \\
  \dot{E}_{plas} &= \dot{E}_{plas,B} + \dot{E}_{plas,nB}, \\
  \dot{E}_{plas,B} &= N_A \sum_i \Delta E_i r_i \quad (\text{Bolsig reactions}), \\
  \dot{E}_{plas,NB} &= -\sum_{i=1}^{N_{sp}} h_i^0 \dot{\omega}_i \quad (\text{non-Bolsig reactions}),
\end{align} 
In the heat source terms, $h_i^0$ is the formation enthalpy, $\Delta E_i$ the threshold energy, and $r_i$ the rate of progress of the $i$-th electron-impact reaction. $\dot{E}_{plas,B}$ and $\dot{E}_{plas,nB}$ are the energy deposited by the plasma for Bolsig and non-Bolsig reactions, respectively. Note that energy stored in excited states by Bolsig reactions is tracked in $\dot{Q}_{react}$ using the ground-to-excited state enthalpy difference. Therefore, nonequilibrium effects are captured while preserving overall energy conservation.

\subsubsection{Gap phase} 
In NRPDs, significant plasma chemistry and energy transfer still continue after the pulse phase ends. During the gap phase, the electron density rapidly decays, while long‑lived ions, radicals, and metastables continue to evolve through chemical pathways and diffuse with the gas flow. The after‑pulse stage (i.e., gap phase) strongly influence ignition enhancement by radical production from plasma chemistry. Moreover, an explicit afterglow during the gap phase is required to capture electron–ion and ion–ion recombination, ambipolar diffusion, and post‑discharge reaction kinetics that cannot be represented by the pulsed discharge alone. Furthermore, including the electrostatic potential through Poisson's equation significantly increases the numerical stiffness due to the short characteristic response time scales of electrons, requiring time steps on the order of a few picoseconds during the pulse phase. During the gap phase, the electric field applied by the anode can be neglected due to its low magnitude, leaving only the ambipolar electric field generated by the residual space charge resulting from the difference between the electron and ion diffusion coefficients. Therfore, Poisson's equation for the electrostatic potential is no longer required. In addition, electrons hardly gain energy from the electric field after pulse phase, and any energy they acquire is rapidly dissipated through frequent collisions with neutral species. So, we assume that electron temperature ($T_e$) equals the gas temperature ($T_g$) during the gap phase. Under these assumptions, the governing equations reduce to the following set of quasi-neutral plasma–flow equations with ambipolar diffusion:
\begin{align}
 \frac{\partial \rho}{\partial t} + \nabla \cdot \left(\rho \vec{u}\right) &= 0, \\
 \frac{\partial \rho \vec{u}}{\partial t} + \nabla \cdot (\rho \vec{u} \vec{u}) &= -\nabla p + \nabla \cdot \tau,  \\
 \frac{\partial n_e}{\partial t} + \nabla \cdot \left( \frac{1}{m_e} \vec{\Gamma}_{e,amb} + \vec{u} n_e \right) &= N_A \dot{\omega}_e + S_{ph,e}, \\
 \frac{\partial \rho Y_{i}}{\partial t} + \nabla \cdot \left(\vec{\Gamma}_{i,amb} + \rho \vec{u} Y_i \right) &= W_{i} \dot{\omega}_{i} + \frac{W_{i} S_{ph,i}}{N_A}, \\
 \frac{\partial \rho h_s}{\partial t} + \nabla \cdot (\rho h_s \vec{u} + \vec{q}) + \tau : \nabla \vec{u} &= \frac{Dp}{Dt} + \dot{Q}_{react} + 
 \frac{3}{2} n_e k_B \left(T_e - T_g\right) \frac{2 m_e}{m_g} \nu +\dot{E}_{plas,nB}.
\end{align} 
where $\vec{\Gamma}_{e, amb}$ and $\vec{\Gamma}_{i, amb}$ are the ambipolar diffusion fluxes of electrons and ion species $i$, respectively, which can be expressed as~\cite{prager2007modeling}:
\begin{align} 
 \vec{\Gamma}_{i,amb} &= - \rho D_i \nabla Y_i + \rho \mu_i \vec{E} Y_i - \beta_i \sum_j^{N_{ch}} \frac{Z_j W_i}{Z_i W_j} \vec{\Gamma}_i, \\
 \beta_i &= \frac{Z_i^2 D_i X_i}{\sum_j Z_j^2 D_j X_j}, \quad \vec{\Gamma}_i = \mu_i \vec{E} n_e - D_i \nabla n_e
\end{align} 
where $N_{ch}$ represents the number of charged species, $\beta_i$ the species mole fraction $X_i$ weighted by the diffusion coefficient, $\vec{\Gamma}_i$ the mass diffusion flux of $i$-th species. Neutrality is conserved when the total flux of charged species equals to zero~\cite{franklin2003basic}:
\begin{align}
 \sum_i^{N_{ch}} \frac{Z_i e \overline{W}}{W_i} \vec{\Gamma}_i &= 0.
\end{align}
where $\overline{W}$ the mean molar mass of the gas mixture. 

\subsection{Photoionization in air} 
Photoionization occurs when UV photons emitted from regions of high electric field in the streamer head ionize neutral molecules ahead of the discharge front. This process generates seed electrons in front of the streamer, enabling its continued propagation through secondary avalanches that sustain the advance of streamers. The underlying reaction is
\begin{align}
 & M + h\nu \rightarrow{} M^+ + e^-,
\end{align} 
where $M$ is a neutral gas molecule and $M^+$ and $e^-$ are the resulting ion and electron. The photon energy is $h\nu$, with $h$ denoting Planck’s constant and $\nu$ the photon frequency. In dry air, the radiative decay of excited \ce{N2} species emits photons. These photons—specifically those in the 12.1 to 12.65 eV energy range—subsequently photoionize \ce{O2} molecules~\cite{bouwman20223d}. According to Zhelezniak~\textit{et al.}~\cite{zhelezniak1982photoionization}, the photoionization rate at the point of observation $\vec{r}$ due to UV photons emitted from source points $\vec{r}'$ is expressed as:
\begin{align}
S_{\text{ph}}(\vec{r}) &= \iiint_{V'} \frac{I(\vec{r}') \, g(R)}{4\pi R^2} \, \mathrm{d}V'.
\end{align}
where $R=|\vec{r}-\vec{r}'|$, $I(\vec{r}')$ the local photon‑emission source term, $g(R)$ the probability that a photon emitted at one point will travel a distance $R$, and $V'$ the entire discharge volume. The emission source term $I(\vec{r}')$ is given by~\cite{luque2007photoionization}
\begin{align}
 I(\vec{r}') &= \frac{p_q}{p+p_q} \zeta \frac{\nu_u}{\nu_i} S_i(\vec{r}). 
\end{align} 
where $p_q$ denotes the quenching pressure, $p_q/(p+p_q)$ the quenching for excited \ce{N2} molecules due to collisions, $\zeta$ the photoionization efficiency, $\nu_u$ the electron impact excitation frequency for level $u$, $\nu_i$ the first Townsend ionization coefficient, and $S_i(\vec{r})$ the ionization rate of \ce{N2}. The function $g(R)$ is 
\begin{align}
 \frac{g(R)}{p_{\ce{O2}}} &= \frac{\text{exp}(-\chi_{min}p_{\ce{O2}}R)-\text{exp}(-\chi_{max}p_{\ce{O2}}R)}{p_{\ce{O2}}R\;\text{ln}(\chi_{max}/\chi_{min})}    
\end{align} 
where $\chi_{min}, \chi_{max}$ are the minimum and maximum absorption coefficients, and $p_{\ce{O2}}$ is the partial pressures \ce{O2}. A novel method was introduced that replaces the integral evaluation in the traditional photoionization model with the solution of Helmholtz equations~\cite{luque2007photoionization}: 
\begin{align}
 \nabla^2 S_{\text{ph}}^j(\vec{r}) - \left(\lambda_j p_{\ce{O2}}\right)^2 S_{\text{ph}}^j(\vec{r}) &= - A_j p_{\ce{O2}} I(\vec{r}).  
\end{align}
where the photoionization source term $S_{\text{ph}}(\vec{r})$ can be written as
\begin{align}
 S_{\text{ph}}^j(\vec{r}) &= \sum_j S_{\text{ph}}^j(\vec{r}),
\end{align}
with individual contributions
\begin{align}
 S_{\text{ph}}^j(\vec{r}) &= \iiint_{V'} \frac{I(\vec{r}') \, A_j p^2_{\ce{O2}} \text{exp}\left(-\lambda_j p_{\ce{O2}} R\right)}{4\pi R^2} \, \mathrm{d}V',
\end{align}
Note that $\lambda_j$ and $A_j$ are coefficients that depend on gas mixtures. The details of code implementation, code verification, and validation are included in \autoref{appendix}.

\subsection{Adaptive time‑step strategy} 
When multiple pulses are applied across the pulse and gap phases, the solver enforces several time‑step constraints to ensure stable and accurate integration of the rapidly changing discharge dynamics. The presence of the Poisson's equation during the plasma discharge makes the system significantly stiffer, to the point that time intervals on the order of picoseconds are required to resolve the fast electrostatic dynamics. Moreover, transitions from the pulse phase to the gap phase, and from the gap phase back to the pulse phase, present a significant challenge, as these sudden phase changes can trigger severe numerical instabilities. To overcome these issues, appropriate time‑step constraints and supplementary stabilization mechanisms are required to prevent numerical instabilities. Here, we outline the adaptive time‑step criteria employed in the solver and describe how each constraint contributes to maintaining stability across both phases. 

The Courant–Friedrichs–Lewy (CFL) condition limits the time step based on the characteristic transport speeds in the system. In particular, the CFL number ensures that advective information does not travel more than one cell length during a single time step, thereby preventing numerical instabilities associated with the discretization of the continuity and momentum equations. For a given cell size $\Delta x$ and maximum characteristic velocity $u_{\text{max}}$, the time step $\Delta t_{CFL}$ must satisfy:
\begin{align}
 & \Delta t_{CFL} \leq \frac{CFL \Delta x}{|u_{\text{max}}|},
\end{align}
where $CFL$ is a user‑defined stability. 

In weakly ionized plasmas, the electron drift velocity and diffusion velocity can be orders of magnitude larger than the bulk gas velocity. Even when the CFL condition based on the fluid velocity is satisfied, the numerical scheme can still become unstable if electrons drift across multiple cells within a single time step. To prevent this, The drift dynamic time step $\Delta t_c$ and diffusion time step $\Delta t_d$ are imposed~\cite{zhu2017nanosecond}:
\begin{align}
 & \Delta t_c = \xi_c \text{min}\left(\frac{\Delta x}{\mu_e D_e}\right), \\
 & \Delta t_d = \xi_d \text{min}\left(\frac{(\Delta x)^2}{D_e}\right),
\end{align}
where $\xi_c$ and $\xi_d$ are the drift and diffusion coefficients, respectively. Both are set to 0.1 during the streamer‑propagation phase, and $\xi_d$ is increased to 0.3 during the gap‑discharge phase. 

The dielectric relaxation time step $\Delta t_{diel}$ is limited by the characteristic time required for charge imbalance to relax through conduction in the plasma:
\begin{align}
 & \Delta t_{diel} = \frac{\epsilon_0}{e n_e \mu_e},
\end{align}
Since the dielectric relaxation time step is most restrictive -- on the order of $ 10^{-15}$ s -- we focus on streamer propagation and glow discharge, for which the dielectric‑relaxation constraint is not utilized in our simulations.

The chemical time‑step constraint arises from the stiffness of the ionization, attachment, recombination, and excitation reactions. In regions of high electron density or strong electric field, the characteristic chemical timescale can become extremely small, and explicit integration becomes unstable unless the time step resolves the fastest reaction rate. The chemical time step $\Delta t_{chem}$ is therefore defined as
\begin{align}
 & \Delta t_{chem} = 0.1 \ \text{min}\left( \Delta t_{CFL}, \Delta t_c, \Delta t_d \right),
\end{align}
This ensures that the source terms in the continuity equations do not change the species densities by more than a small fraction within a single time step.

\subsection{Plasma chemistry parser in OpenFOAM} 
The OpenFOAM-6 provides two primary chemistry readers: the standard CHEMKIN reader and the native FOAM-format chemistry reader (foamChemistryReader). The built-in CHEMKIN reader is capable of importing detailed gas-phase kinetic mechanisms, including thermodynamic data in the standard CHEMKIN/NASA-7 polynomial format. However, the following limitations exist when using the current native reader to plasma simulations:
\begin{itemize}
    \item The current chemistry readers are limited to neutral species and their associated chemical reactions. Plasma chemistry—including electron-impact dissociation, excitation, ionization, recombination, and ion–molecule reactions—is not recognized by the default parser and therefore cannot be implemented directly.
    \item Reaction rate coefficients for Bolsig reactions depend on the cross section $\sigma(\vec{E})$, the electron speed $v_e$, and the electron energy distribution function (EEDF), the electron energy distribution function (EEDF); these cannot be accurately represented by a standard Arrhenius form.
    \item The EEDF can either be assumed Maxwellian or computed as a non-Maxwellian solution using a Boltzmann solver. The latter is computationally intensive in three dimensions, particularly for mixtures involving many species and complex collision processes.
\end{itemize}
Consequently, the default framework is insufficient for simulations of plasma-assisted combustion (PAC), necessitating the development of an extended chemistry reader capable of incorporating charged species and plasma-specific reaction pathways. To overcome these limitations, a custom chemistry reader, \textit{chemkinPlasmaReader}, is developed, which extends the standard OpenFOAM chemistry reader (\textit{chemkinReader}) to fully support charged species, plasma chemistry, and their associated transport and reaction dynamics.

\subsubsection{\textit{chemkinPlasmaReader}}
The standard \textit{chemkinReader} is restricted to neutral species and conventional combustion chemistry because it assumes that all reaction rate coefficients strictly follow the Arrhenius form. In contrast, the \textit{chemkinPlasmaReader} expands the species set to include charged and electronically excited species, enabling support for conventional combustion chemistry as well as electron‑impact, dissociation, excitation, ionization, recombination, and ion–molecule reactions. Furthermore, the \textit{chemkinPlasmaReader} is designed to parse an extended CHEMKIN‑style input format for plasma chemistry, in which each reaction type is associated with a dedicated parsing rule and mapped to a corresponding plasma‑reaction class. This design enables consistent handling of heterogeneous plasma kinetics, allowing the reader to retrieve polynomial rate‑coefficient expressions as functions of electron temperature and gas temperature, in addition to standard Arrhenius rates. The parsed plasma reactions are then assembled into the chemistry model alongside conventional combustion reactions. 

\begin{figure}
    \centering
    \includegraphics[width=0.8\linewidth]{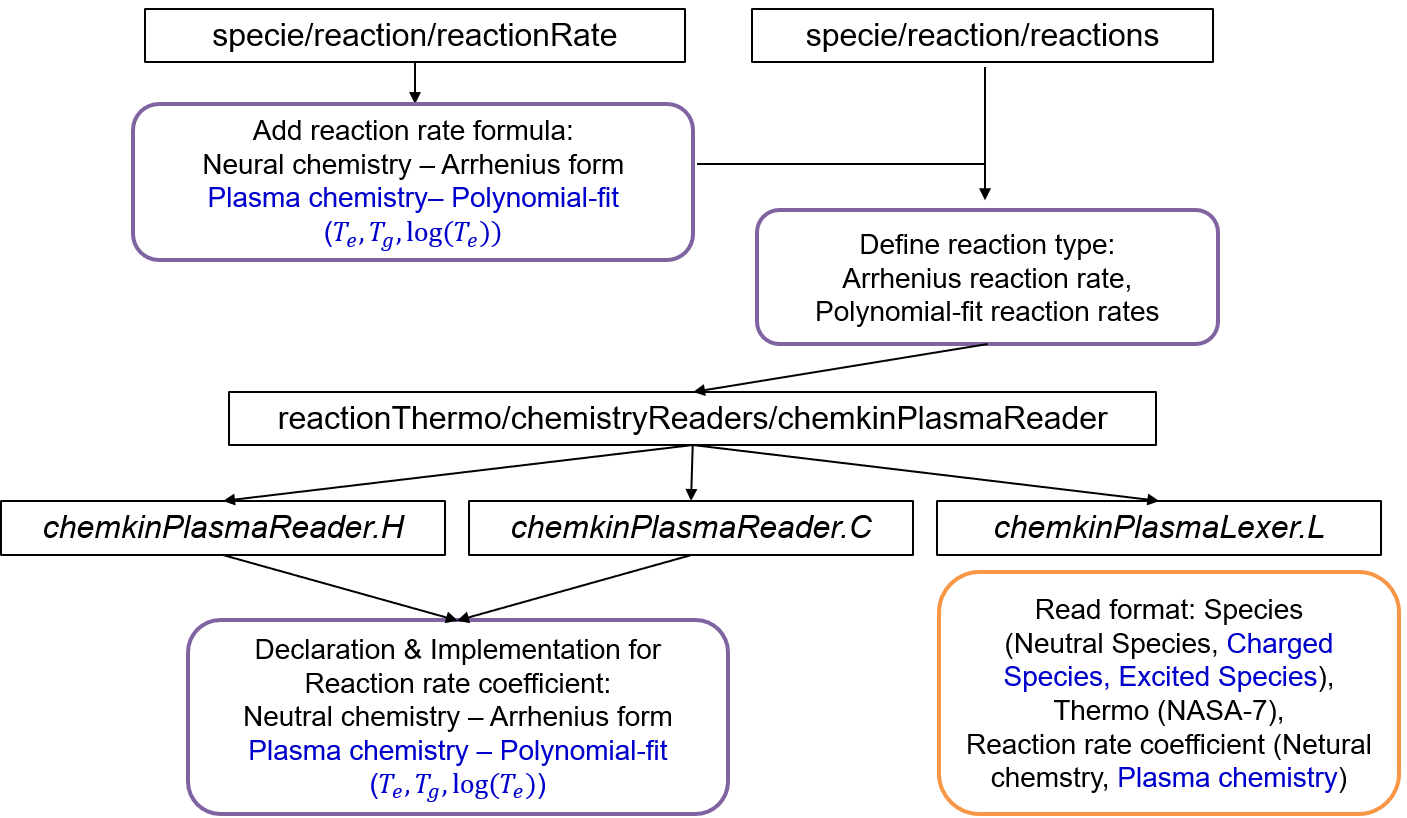}
    \caption{A schematic diagram illustrating the integration of plasma and combustion chemistry into the \textit{chemkinPlasmaReader} framework in OpenFOAM.}
    \label{ChemkinReaderDiagram}
\end{figure}

Figure~\ref{ChemkinReaderDiagram} illustrates the schematic structure of \textit{chemkinPlasmaReader}, which extends the conventional CHEMKIN-based chemistry parsing framework to incorporate plasma reactions. The overall architecture is divided into two primary branches: (i) reaction-rate formulation (\textit{specie/reaction/reactionRate} in \textit{src}) and (ii) reaction definition (\textit{specie/reaction/reactions} in \textit{src}). The \textit{reactionRate} module supports both the conventional Arrhenius formulation for neutral chemistry and a polynomial-fitting approach for plasma chemistry, the latter enabling rate expressions as functions of electron temperature, gas temperature, and the logarithm of electron temperature. The reactions module assigns appropriate rate models to each reaction, allowing Arrhenius and polynomial-fit reaction types to coexist within a unified framework. These components are interfaced through the \textit{chemkinPlasmaReader}, which comprises \texttt{chemkinPlasmaReader.H}, \texttt{chemkinPlasmaReader.C}, and \texttt{chemkinPlasmaLexer.L} for declaration, implementation, and input parsing. The readers are responsible for parsing CHEMKIN-format input files, including species (neutral, charged, and electronically excited species), thermodynamic data based on NASA 7-coefficient polynomials, and reaction mechanisms. In particular, the lexer is extended to recognize plasma reaction entries, while the reader constructs corresponding reaction objects with appropriate rate formulations. This modular design enables integration of plasma-assisted chemistry into existing combustion solvers while maintaining compatibility with standard CHEMKIN workflows.

\subsubsection{Plasma reaction kinetics, reaction rate, and transport coefficients}
\begin{figure}
    \centering
    \includegraphics[width=\linewidth]{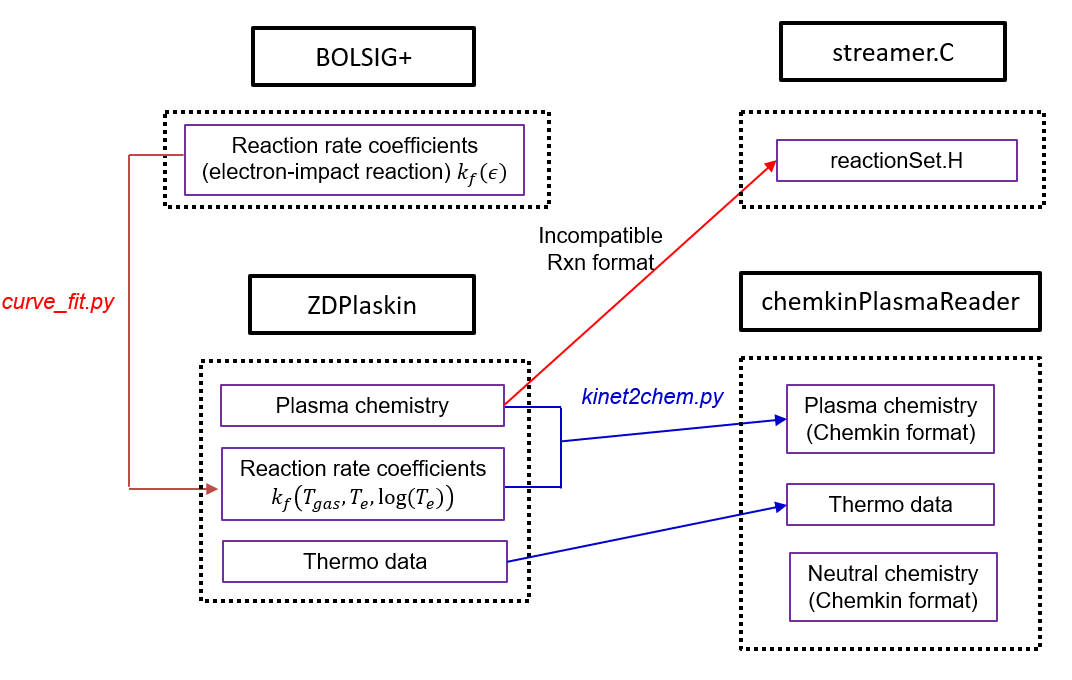}
    \caption{An overview for generating plasma chemistry input files in Chemkin format. Reaction rate coefficients for electron-impact reactions from BOLSIG+~\cite{hagelaar2016brief} are fitted and incorporated into ZDPlaskin chemistry format (\texttt{kinet.inp}), and subsequently converted to the Chemkin format.}
    \label{Kinetics}
\end{figure}

Low‑temperature plasmas represent highly nonequilibrium systems in which electrons, ions, and neutral species often exhibit distinct characteristic temperatures, making reaction rates strongly dependent on the local thermodynamic state. In particular, electron‑impact reactions must be treated independently of the bulk gas temperature, as their rate coefficients depend primarily on the electron energy distribution function (EEDF). To incorporate this without the prohibitive cost of solving the Boltzmann equation during runtime, we employ a pre‑tabulated kinetic‑processing pipeline, as illustrated in Figure~\ref{Kinetics}. Reaction rate coefficients for electron-impact reactions $k_f$ are first generated using offline  BOLSIG+~\cite{hagelaar2016brief} as functions of the electron mean energy ($\varepsilon$). The rate coefficients are then processed using a custom \texttt{curve\_fit.py} utility to produce functional expressions dependent on the electron temperature $T_e$, or $\log(T_e)$ in accordance with ZDPlaskin formatting. These data are then converted into a unified CHEMKIN format using \texttt{kinet2chem.py}. Any reactions incompatible with the CHEMKIN parser—primarily vibrational quenching reactions—are isolated in \texttt{reactionSet.H} for direct evaluation within the solver’s source code. At runtime, the \textit{chemkinPlasmaReader} parses the complete mechanism, including neutral, charged, and electronically excited species, NASA‑7 thermodynamic polynomials, and multi‑parameter reaction‑rate expressions. Each reaction is mapped to a dedicated class that evaluates its rate based on the instantaneous local plasma state, enabling consistent and efficient computation of plasma‑assisted reaction source terms. The transport coefficients of electron and mean electron energy are obtained from BOLSIG+~\cite{hagelaar2016brief}, where they are tabulated as functions of the local mean electron energy per electron $\varepsilon$. The local value of $\varepsilon$ is subsequently used in a linear interpolation to obtain the corresponding coefficients within our OpenFOAM framework, updating them at each iteration. As a result, the transport and reaction coefficients remain self‑consistent with the evolving plasma state without requiring the explicit solution of the electron energy distribution function

\section{Results and discussion} \label{s_elongation}
In this section, we investigate plasma morphology within a 3D channel featuring a pin-to-pin electrode configuration in atmospheric pressure air under nanosecond repetitive pulsed discharges (NRPDs). Two distinct simulation cases are considered, featuring a low-velocity regime (40 m/s) and a high-velocity regime (100 m/s) for the inlet cross-flow. Specifically, the main focus is on how the cross-flow velocity governs plasma morphology at the midpoint between the electrodes and near the anode, which significantly affects thermal accumulation and the formation of a continuous high-temperature channel downstream. Moreover, we examine the physical mechanism behind the snap-back event, where the formation of a new low-resistance conduction path abruptly redirects current and collapses the elongated plasma back toward the shortest inter-electrode path.

\begin{figure}
    \centering
    \begin{subfigure}{0.45\linewidth}
        \centering
        \includegraphics[width=\textwidth,trim=5 5 5 5,clip]{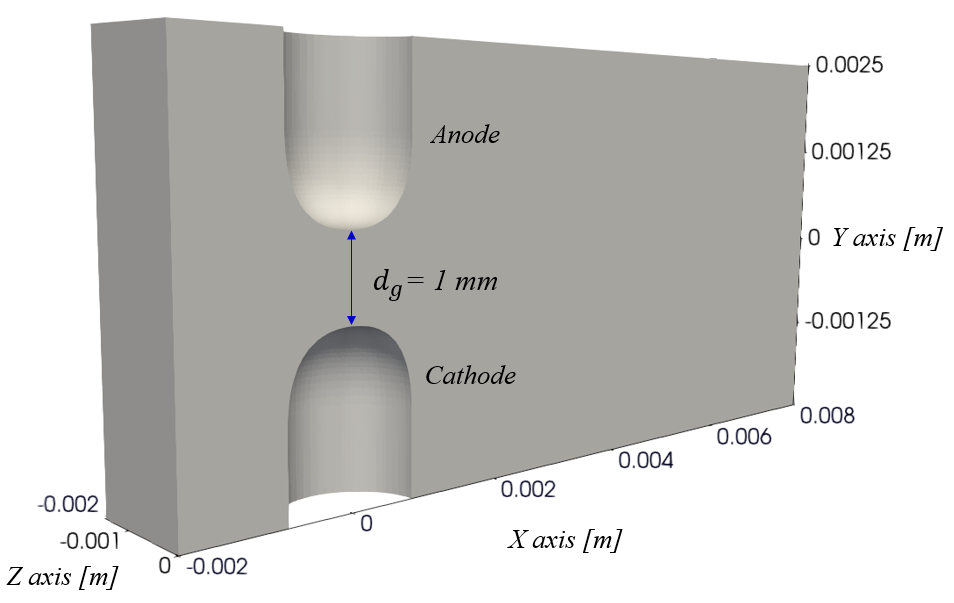}
        \caption{}
    \end{subfigure}
    \begin{subfigure}{0.54\linewidth}
        \centering
        \includegraphics[width=\textwidth,trim=5 5 5 5,clip]{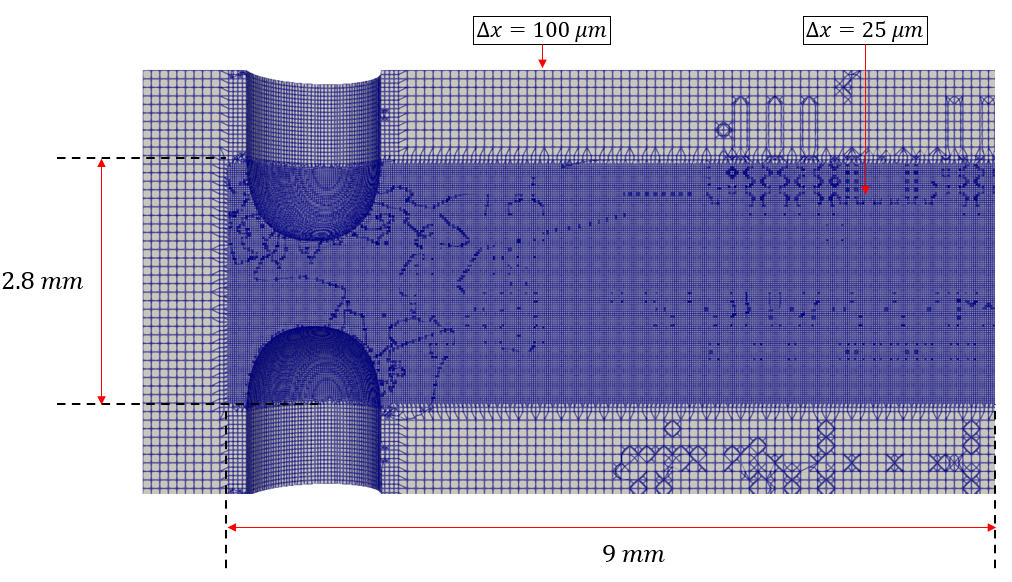}
        \caption{}
    \end{subfigure}
    \caption{Computational domain and mesh characteristics: (a) 3D view of the pin-to-pin electrode configuration with an inter-electrode gap $d_g = 1$ mm, (b) Side view of the refined hexahedral mesh generated via snappyHexMesh, highlighting the baseline resolution ($\Delta x = 100\ \mu\text{m}$) and the two-level refinement zones ($\Delta x = 25\ \mu\text{m}$) in the discharge and downstream regions.}
    \label{3D_channel}
\end{figure}

\subsection{Computational setup} 
Figure~\ref{3D_channel} illustrates the 3D domain featuring a pin-to-pin electrode configuration. The domain size is 10 mm $\times$ 5 mm $\times$ 4 mm, with the electrodes positioned at the center of the $x$-axis. The electrode diameter $d$ is 1.57 mm, and the inter-electrode gap $d_g$ is set to 1 mm. The mesh is generated using \texttt{snappyHexMesh}, a utility within the OpenFOAM framework designed to refine a background hexahedral mesh and conform it to complex 3D geometries defined by STL surface files, thereby generating a hexahedral-dominant unstructured mesh with surface refinement suitable for resolving complex geometries. A baseline cell size of $\Delta x = 100\ \mu\text{m}$ is employed throughout the bulk flow region, while two levels of mesh refinement are applied within a specified region ($9~\text{mm} \times 2.8~\text{mm} \times 1.6~\text{mm}$), resulting in a local resolution of $\Delta x = 25\ \mu\text{m}$ to accurately resolve the plasma between the two electrodes and in the downstream region. Two boundary layers with an expansion ratio of 1.1 are extruded from the electrode surfaces to better resolve near-wall effects, resulting in a total of 2{,}605{,}817 cells. All schemes are selected based on considerations of numerical stability: A first-order Euler scheme is employed for temporal discretization. The gradient terms are discretized using the Gauss linear scheme, while the divergence terms for the convection of the flow, species, electron density, and electron energy are discretized using the first-order upwind scheme. The viscous stress term is discretized using the Gauss linear scheme. Diffusion terms are evaluated using the Gauss linear corrected scheme, and linear interpolation is used for face-centered quantities. A corrected scheme is employed for the surface-normal gradients to improve the accuracy of the discretization on non-orthogonal meshes. Note that Poisson's equation for the electrostatic potential and the Helmholtz equations for photoionization are solved using the GAMG (Geometric-Algebraic Multi-Grid) solver, which is well suited for solving large sparse linear systems arising from elliptic equations. Table~\ref{tab:airChemistry} summarizes the species employed in the simulations. A total of 32 species and 234 reactions are incorporated to represent the plasma discharge in air, including plasma and combustion chemistry~\cite{taneja20241d, zhong2023understanding}, which have been well validated for \ce{NH3}/air mixtures using ZDPlaskin~\cite{pancheshnyi2008zdplaskin} and in 1D streamer simulations. Note that \ce{NH3} and its associated sub-species are removed from the mechanism. Note that all numerical simulations are conducted on the Raptor HPC cluster where each compute node features dual AMD EPYC 7713 64-Core Processors (128 physical cores per node, Zen 3 microarchitecture). High-performance parallel execution are partitioned across 1,568 processors. For the $40\text{ m/s}$ flow speed case, total simulation time was 240 wall-clock hours, whereas the $100\text{ m/s}$ case required 192 wall-clock hours.

\begin{table}[h!]
\centering
\caption{Species included in the air plasma‑chemistry mechanism, categorized into neutral, vibrationally excited, electronically excited, and charged species.} \label{tab:airChemistry}
\begin{tabularx}{0.95\textwidth}{lX}
\hline
\textbf{Type} & \textbf{Species} \\
\hline
Neutrals &
\ce{N} \; \ce{N2} \; \ce{O} \; \ce{O2} \; \ce{O3} \; \ce{NO} \; \ce{NO2} \; \ce{N2O} \; \ce{Ar}\\
Vibrationally excited neturals & 
\ce{N2(V1)} \; \ce{N2(V2)} \; \ce{N2(V3)} \; \ce{N2(V4)} \; \ce{N2(V5)} \; \ce{N2(V6)} \; \ce{O2(V1)} \; \ce{O2(V2)} \;
\ce{O2(V3)} \; \ce{O2(V4)} \\
Excited neutrals &
\ce{N2(A)} \; \ce{N2(B)} \; \ce{N2(C)} \; \ce{N2(a)} \; \ce{N(^2D)} \; \ce{O2(A)} \; \ce{O(^1D)} \; \ce{O(^1S)} \\
Cations &
\ce{N2+} \; \ce{O2+}
\\
Anions/electron &
\ce{O2-} \; \ce{O-} \; \ce{e-}
\\
\hline
\end{tabularx}
\end{table}

\subsection{Pulse configuration}
\begin{figure}
    \centering
    \includegraphics[width=\linewidth]{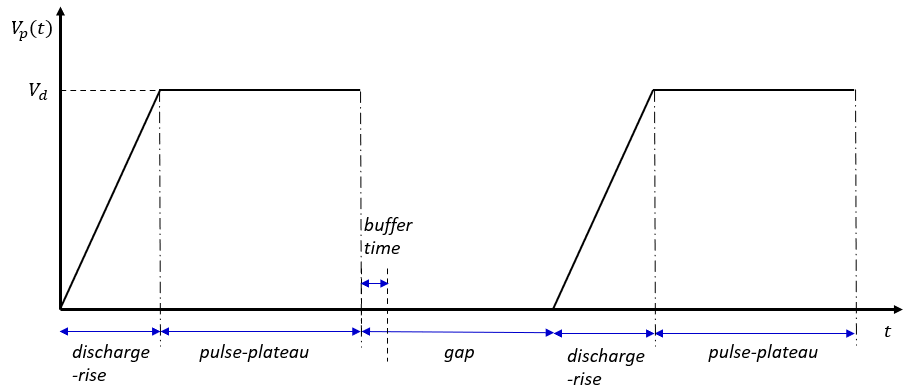}
    \caption{Schematic representation of the three-stage pulse configuration: the discharge-rise phase, the pulse-plateau phase, and the gap phase. The buffer time is highlighted as the transition interval designed to maintain numerical stability and prevent time-step overshooting at the onset of the gap phase.}
    \label{pulse_configuration}
\end{figure}

The pulse profile is critical for maintaining numerical stability. Abruptly applying the full peak voltage at the beginning of the pulse can produce strong electric fields and rapid changes in the coupled plasma variables, potentially requiring smaller time steps for numerical stability and temporal accuracy. To address this computational challenge, three stages are implemented in our simulations: the discharge-rise, pulse-plateau, and gap phases. Figure~\ref{pulse_configuration} shows the multi-pulse configuration employed in our solver. In the discharge-rise stage, the voltage increases linearly to the peak applied voltage, $V_d$. Once it reaches $V_d$, the constant voltage is maintained in the pulse-plateau stage. Following the pulse-plateau phase, a buffer time is introduced to ensure that the increased time step immediately after the pulse does not compromise numerical stability. In our simulations, a peak applied voltage ($V_d$) of 3.5 kV is employed with a total pulse width of 7 ns, consisting of a 1 ns discharge-rise phase followed by a 6 ns pulse-plateau phase. A pulse repetition frequency (PRF) of $50\ \text{kHz}$ is used, with a 3 ns buffer time introduced to stabilize the transition between the pulse-plateau and gap phases.

\begin{figure}[b!]
    \centering
    \begin{subfigure}{0.17\linewidth}
        \centering
        \includegraphics[width=\textwidth,trim=5 5 5 5,clip]{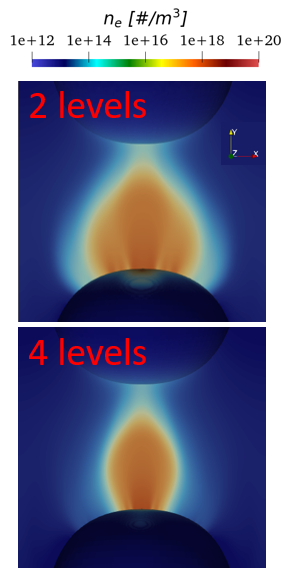} 
        \caption{} \label{14a}
    \end{subfigure}
    \begin{subfigure}{0.4\linewidth}
        \centering
        \includegraphics[width=\textwidth,trim=5 5 5 5,clip]{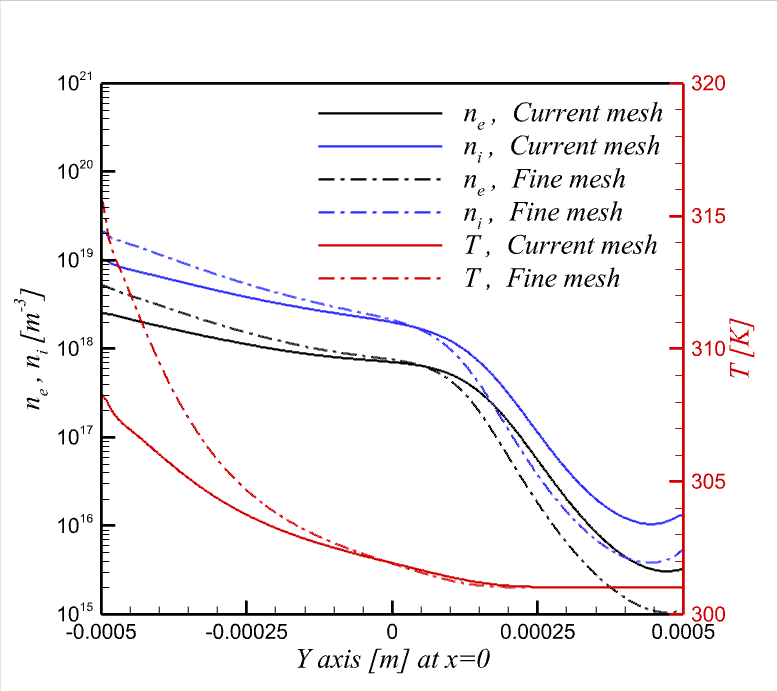}
        \caption{} \label{14b}
    \end{subfigure}
    \begin{subfigure}{0.4\linewidth}
        \centering
        \includegraphics[width=\textwidth,trim=5 5 5 5,clip]{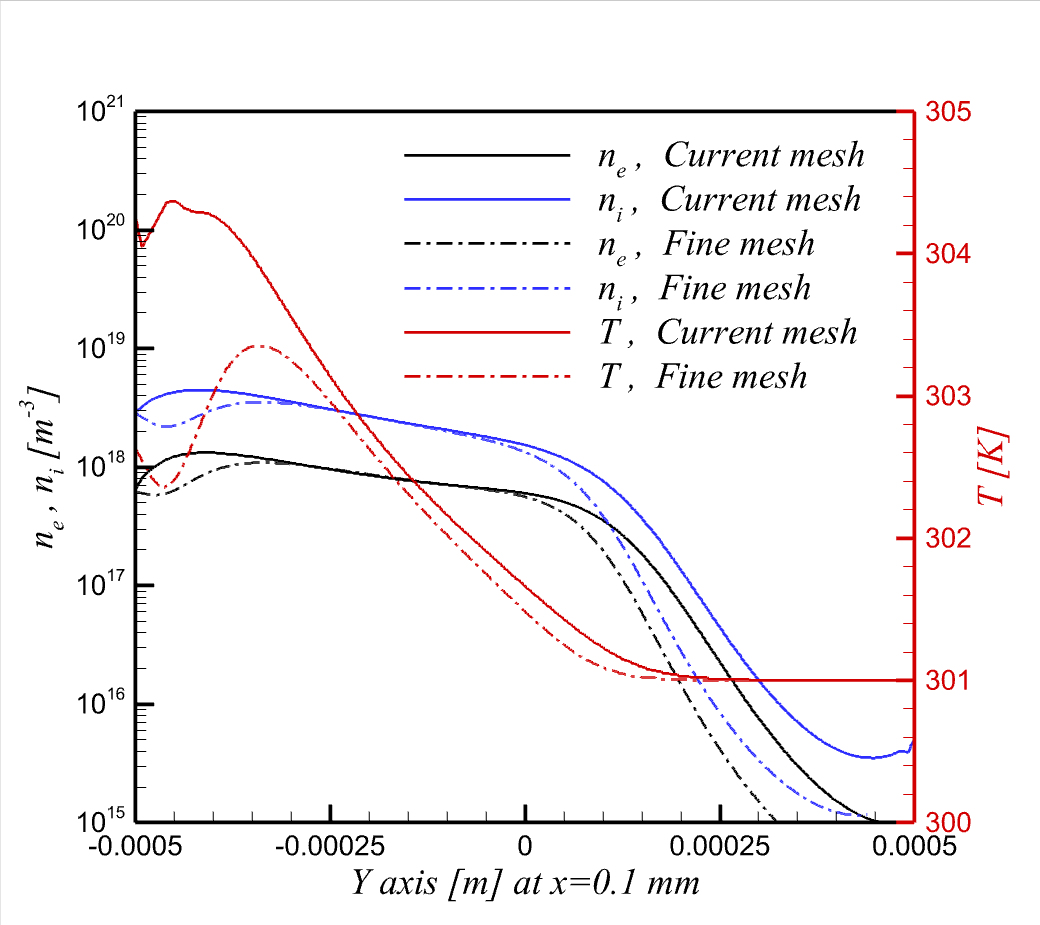}
        \caption{} \label{14c}
    \end{subfigure}
    \caption{Grid resolution test at $t = 10\ \text{ns}$: 
    (a) Spatial distribution of electron density ($n_e$) comparing 2 levels of mesh refinement (top) against 4 levels of mesh refinement (bottom), 
    (b) Profiles of electron density ($n_e$), ion density ($n_i$), and neutral gas temperature ($T$) evaluated from cathode ($y=-0.5$ mm) to anode ($y=0.5$ mm) along the central discharge axis ($x = 0\text{ mm}$), 
    (c) Corresponding profiles evaluated along the off-axis slice ($x = 0.1\text{ mm}$).}
    \label{grid_resolution}
\end{figure}

\subsection{Grid resolution and convergence} 
Here, a grid resolution test is performed using the current mesh (two refinement levels, $\Delta x = 25\ \mu\text{m}$) and a finer mesh between the two electrodes to assess the grid independence of the spatial discretization. The fine mesh employs four levels of mesh refinement to achieve a local resolution of $\Delta x = 6.25\ \mu\text{m}$, resulting in a total of 22 million cells, which is sufficient to resolve the photoionization source term at the direct numerical simulation (DNS) level~\cite{eichwald2012effects}. Figure~\ref{14a} illustrates a spatial comparison of the electron density distribution at 10$\ \text{ns}$ between the current and the finer meshes. The current mesh shows a high-density electron region that extends horizontally much farther into the surrounding region near the cathode than that of the finer mesh. To quantitatively evaluate these discrepancies, Figs.~\ref{14b} and \ref{14c} show the localized profiles of electron density ($n_e$), ion density ($n_i$), and neutral gas temperature ($T$) extracted along the centerline ($x=0$) and at an off-axis position ($x=0.1$ mm), respectively. At the center ($x=0$), the current mesh slightly underpredicts both the peak charged particle densities and the maximum gas temperature near the cathode surface ($y = -0.5$ mm). Conversely, moving away from the cathode boundary into the bulk plasma region, the profiles show good agreement between the two mesh configurations at both $x=0$ and $x=0.1$ mm. The current mesh with two levels of refinement provides a computationally efficient yet highly accurate representation of the plasma morphology in the bulk plasma and downstream regions.

\begin{figure}[b!]
    \centering
    \includegraphics[width=\linewidth]{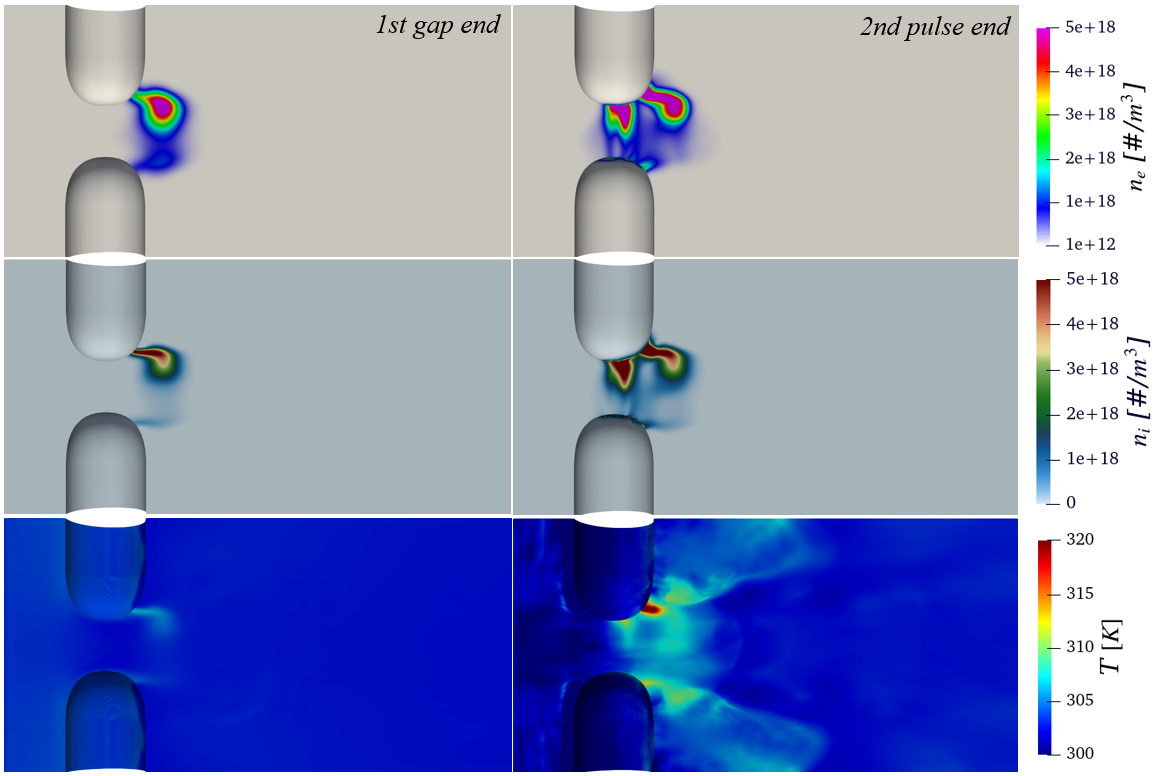}
    \caption{Evolution of electron density ($n_e$), ion density ($n_i$), and gas temperature ($T$) under a 40 m/s cross-flow velocity. The left figures show the states at the end of the first gap phase ($t=20\ \mu\text{s}$), while the right figures show the states after the end of the second pulse phase ($t=21\ \mu\text{s}$).}
    \label{v40_1gap}
\end{figure}

\begin{figure}[t!]
    \centering
    \includegraphics[width=\linewidth]{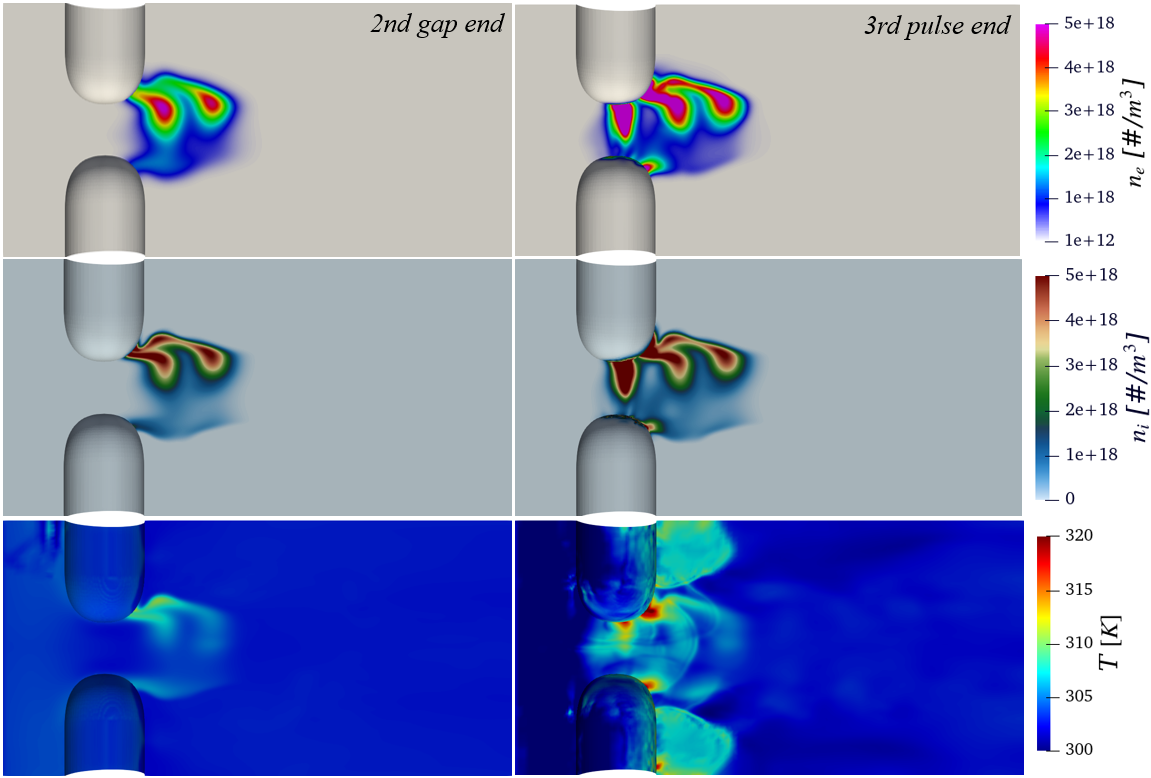}
    \caption{Evolution of electron density ($n_e$), ion density ($n_i$), and gas temperature ($T$) under a 40 m/s cross‑flow velocity. The left figures show the states at the end of the second gap phase ($t=40\ \mu\text{s}$), while the right figures show the states after the end of the third pulse phase ($t=41\ \mu\text{s}$).}
    \label{v40_2gap}
\end{figure}

\begin{figure}[h!]
    \centering
    \includegraphics[width=\linewidth]{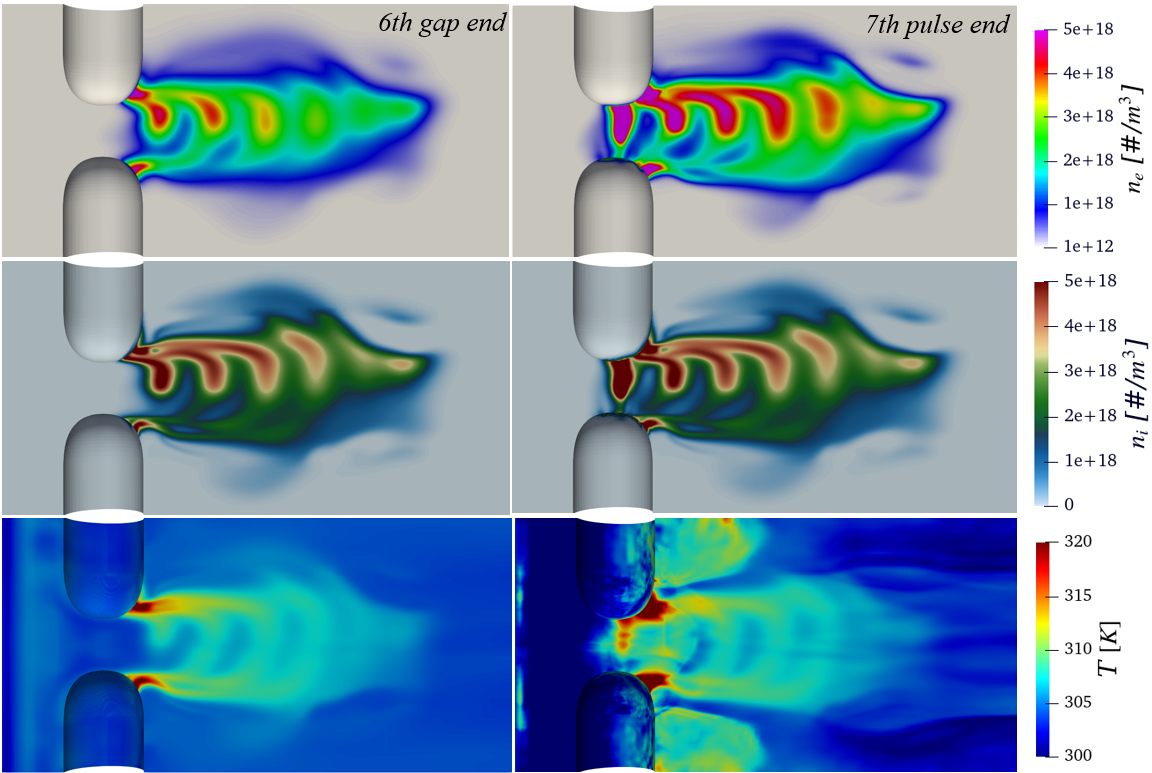}
    \caption{Evolution of electron density ($n_e$), ion density ($n_i$), and gas temperature ($T$) under a 40 m/s cross‑flow velocity. The left figures show the states at the end of the sixth gap phase ($t=120\ \mu\text{s}$), while the right figures show the states after the end of the seventh pulse phase ($t=121\ \mu\text{s}$).}
    \label{v40_6gap}
\end{figure}

\begin{figure}[h!]
    \centering
    \begin{subfigure}{0.49\linewidth}
        \centering
        \includegraphics[width=\textwidth,trim=5 5 5 5,clip]{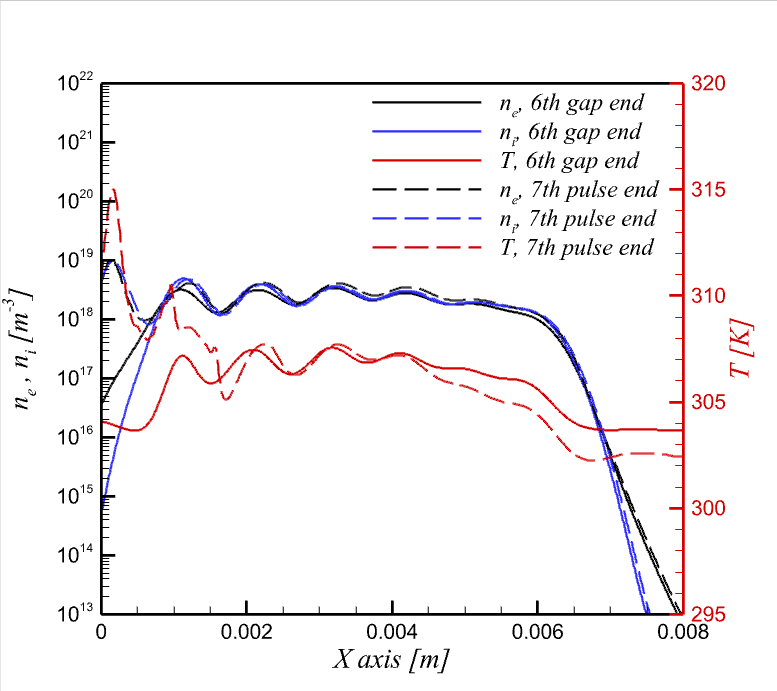}
        \caption{} \label{18a}
    \end{subfigure}
    \begin{subfigure}{0.49\linewidth}
        \centering
        \includegraphics[width=\textwidth,trim=5 5 5 5,clip]{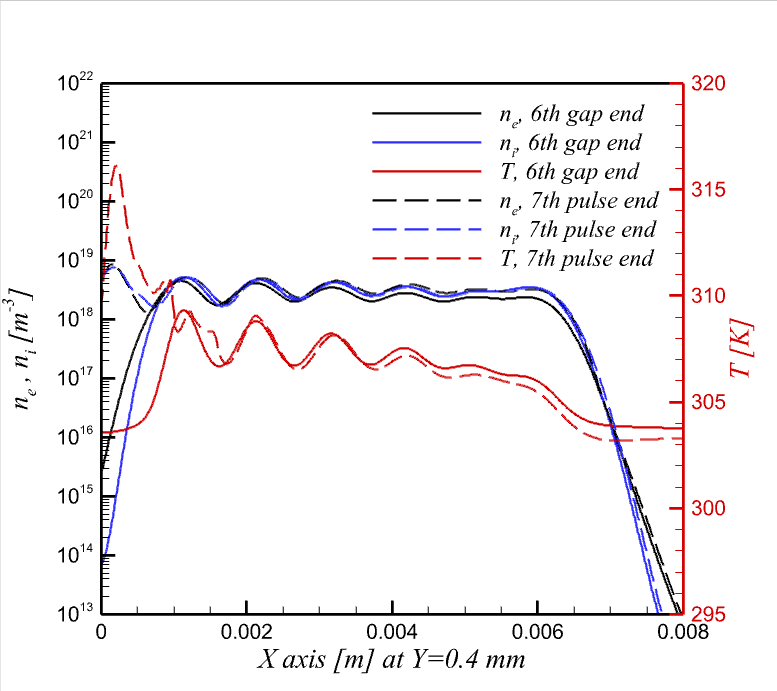}
        \caption{} \label{18b}
    \end{subfigure}
    \caption{Profiles of electron density ($n_e$), ion density ($n_i$), and gas temperature ($T$) along the longitudinal($x$-axis) at the end of the sixth gap ($t=120\ \mu\text{s}$) and the end ofthe seventh pulse ($t=121\ \mu\text{s}$). (a) At the central axis, (b) extracted at $y = 0.4\text{ mm}$. The coordinate system's origin $(0, 0, 0)$ is positioned at the geometric center between the electrode tips.}
    \label{v40_plots}
\end{figure}

\subsection{Effects of cross-flow velocity on plasma morphology} 
The morphological evolution of the plasma is investigated under two different cross‑flow velocities (40 and 100 m/s) at a pulse repetition frequency (PRF) of 50 kHz. Figure~\ref{v40_1gap} showcases the evolution of electron density, ion density, and gas temperature at a cross-flow velocity of 40 m/s, comparing the state at the end of the first gap phase with that at the end of the second pulse. Note that a linear scale is used for both electron and ion densities to observe the plasma morphology clearly. At the end of the first gap phase, the electrons and ions are carried downstream by the cross-flow and primarily accumulate near the anode edge due to strong ionization. After the second pulse, a new conductive channel forms across the electrode gap. Interestingly, an increase in electron and ion densities is observed in the previous channel around the anode and cathode edges. The temperature evolution also confirms that additional energy is deposited into the previously formed channel during the first breakdown, resulting in a higher temperature than that of the newly formed channel after the second pulse. 

Following the subsequent discharge cycle, Figure~\ref{v40_2gap} shows the evolution of the plasma and temperature from the end of the second gap phase to the end of the third pulse phase. The conductive channel created during the first pulse is carried further downstream, thereby expanding the bulk plasma volume. After the third pulse, it is evident that the electron density increases in the previously formed channels, particularly near the anode edge. Accordingly, the strong thermal energy deposited by the plasma into the previous (second) channel contributes to the formation of temperature hotspots near the electrode edges, reaching approximately 320 K. As the discharge progresses through multiple cycles (Figure~\ref{v40_6gap}), the conductive channels exhibit extensive lateral expansion, occupying a significantly wider downstream domain than observed in earlier cycles. An interesting feature of applying multiple pulses is that elevated electron density persists in the previous channel near the anode after the pulse phase. The high electron density promotes continued energy deposition into the previously formed conductive channels, resulting in substantial thermal accumulation and the formation of a continuous high-temperature channel extending across the downstream region, which can contribute to the establishment of stable, extended reaction zones. Figure~\ref{v40_plots} illustrates the one-dimensional profiles of electron density, ion density, and gas temperature at the end of each gap and pulse phase along the $x$-axis at $y=0$ (midpoint between the electrodes) and $y=0.4\text{ mm}$ (0.1 mm below the anode). Notably, the ion density remains relatively unchanged between the end of the sixth gap phase and the end of the seventh pulse. In contrast, the seventh pulse induces an increase in electron density at the midpoint (Figure~\ref{18a}), which becomes significantly more pronounced closer to the anode (Figure~\ref{18b}).

\begin{figure}[t!]
    \centering
    \includegraphics[width=\linewidth]{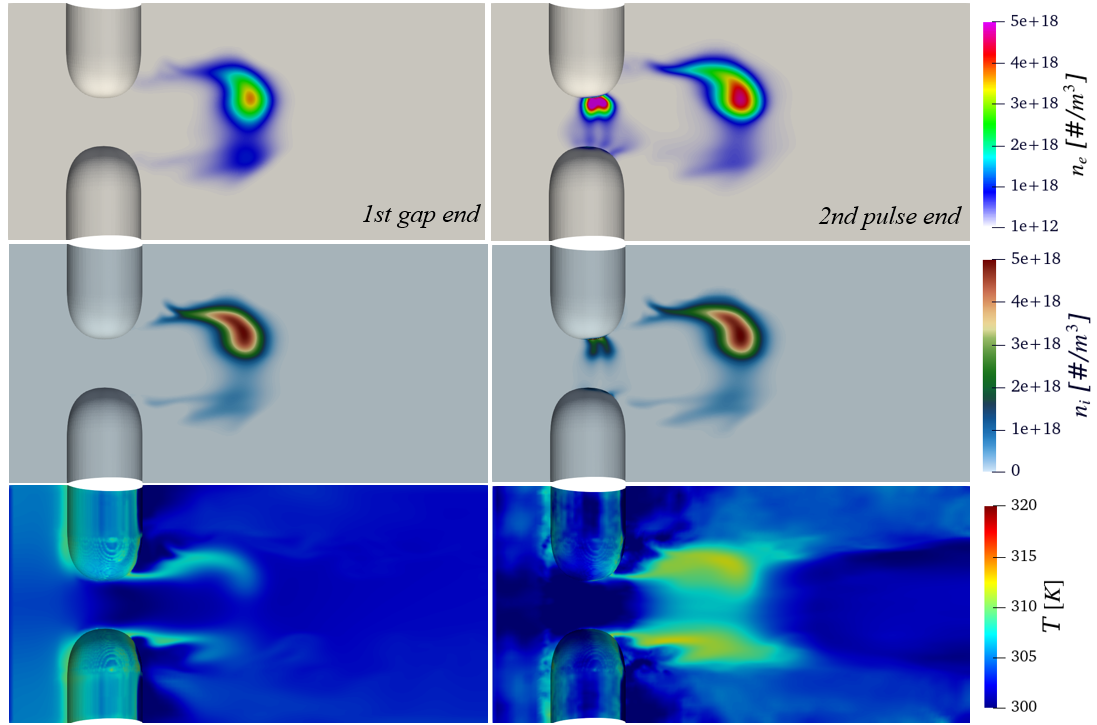}
    \caption{Evolution of electron density ($n_e$), ion density ($n_i$), and gas temperature ($T$) under a 100 m/s cross‑flow velocity. The left figures show the states at the end of the first gap phase ($t=20\ \mu\text{s}$), while the right figures show the states after the end of the second pulse phase ($t=21\ \mu\text{s}$).}
    \label{v100_1gap}
\end{figure}

\begin{figure}[h!]
    \centering
    \includegraphics[width=\linewidth]{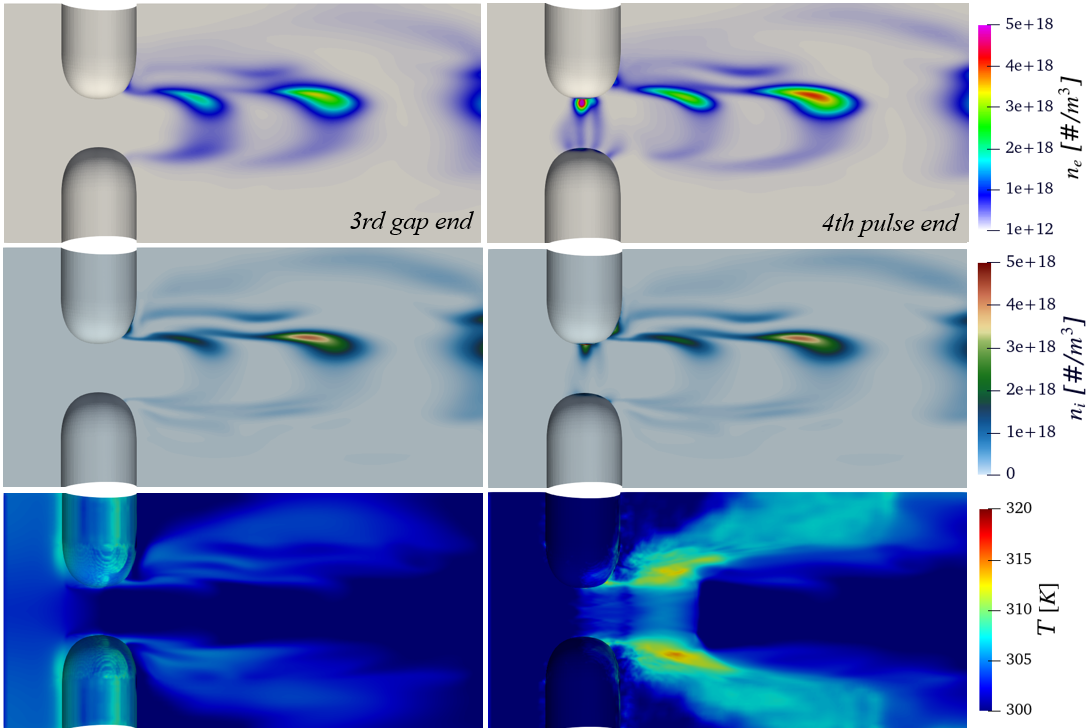}
    \caption{Evolution of electron density ($n_e$), ion density ($n_i$), and gas temperature ($T$) under a 100 m/s cross‑flow velocity. The left figures show the states at the end of the third gap phase ($t=60\ \mu\text{s}$), while the right figures show the states after the end of the fourth pulse phase ($t=61\ \mu\text{s}$).}
    \label{v100_3gap}
\end{figure}

\begin{figure}[h!]
    \centering
    \begin{subfigure}{0.49\linewidth}
        \centering
        \includegraphics[width=\textwidth,trim=5 5 5 5,clip]{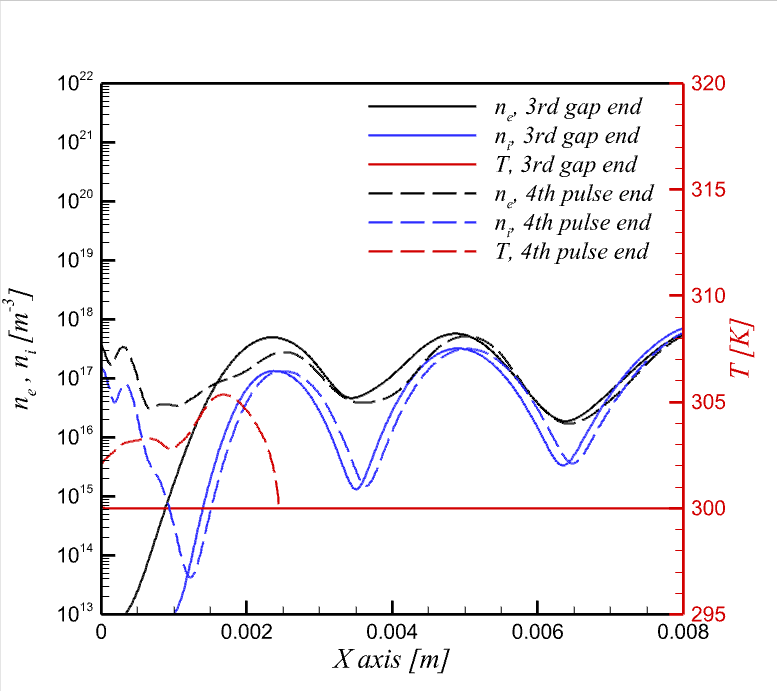}
        \caption{} \label{21a}
    \end{subfigure}
    \begin{subfigure}{0.49\linewidth}
        \centering
        \includegraphics[width=\textwidth,trim=5 5 5 5,clip]{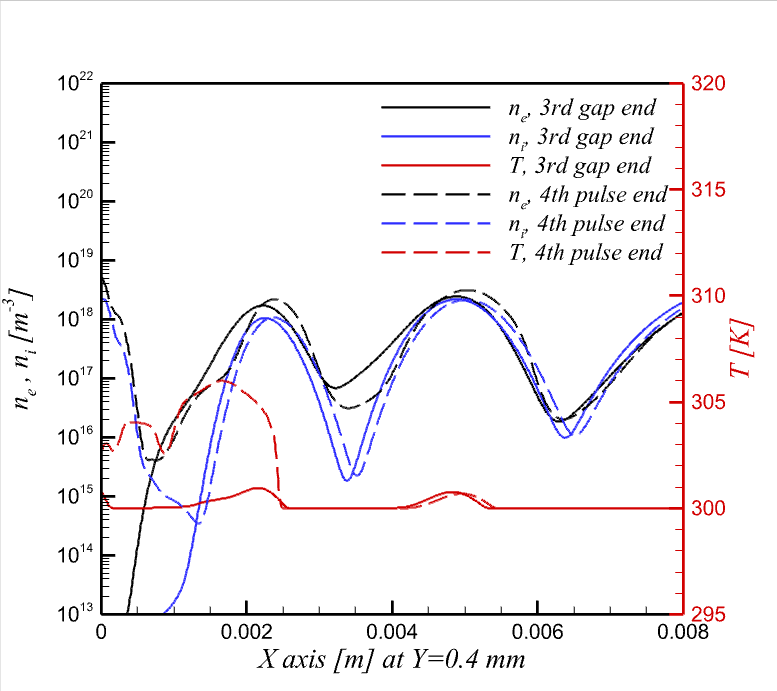}
        \caption{} \label{21b}
    \end{subfigure}
    \caption{Profiles of electron density ($n_e$), ion density ($n_i$), and gas temperature ($T$) along the $x$-axis recorded at the end of the third gap ($t=60\ \mu\text{s}$) and the fourth pulse discharge ($t=61\ \mu\text{s}$): (a) at the central axis, and (b) extracted at $y = 0.4\text{ mm}$. The coordinate system's origin $(0, 0, 0)$ is positioned at the geometric center between the electrode tips.}
    \label{v100_plots}
\end{figure}

We now examine the morphological evolution of the plasma under a high cross-flow velocity at a PRF of 50 kHz. Figure~\ref{v100_1gap} illustrates the evolution of electron density, ion density, and gas temperature at a cross-flow velocity of 100 m/s, comparing the state at the end of the first gap phase with that at the end of the second pulse. By the end of the first gap phase, the electrons and ions are completely advected downstream, and the temperature field shows that the thermal energy is stretched into a thin, split wake flanking the high-speed core flow. At the end of the second pulse, a new conductive channel forms directly between the electrodes, while thermal energy accumulates along the previous channel, forming two distinct high-temperature pathways that flank the central channel. Figure~\ref{v100_3gap} shows the continued evolution of the plasma and temperature from the end of the third gap phase to the end of the fourth pulse. The electrons and ions are advected further downstream compared to those shown in Figure~\ref{v100_1gap}, but the temperature field shows no significant thermal plume between the electrodes following the newly formed conductive channel even after the pulse end. Figure~\ref{v100_plots} illustrates the one-dimensional profiles of electron density, ion density, and gas temperature at the end of each gap and pulse phase along the $x$-axis at $y=0$ (midpoint between the electrodes) and $y=0.4\text{ mm}$ (0.1 mm below the anode). Due to the significant decrease in electron and ion densities between pulses compared with the lower cross-flow velocity of 40 m/s, energy deposition from the previously conductive channels is significantly reduced that the temperature is lower than that at the lower cross-flow velocity.

\begin{figure}[b!]
    \centering
    \includegraphics[width=\linewidth]{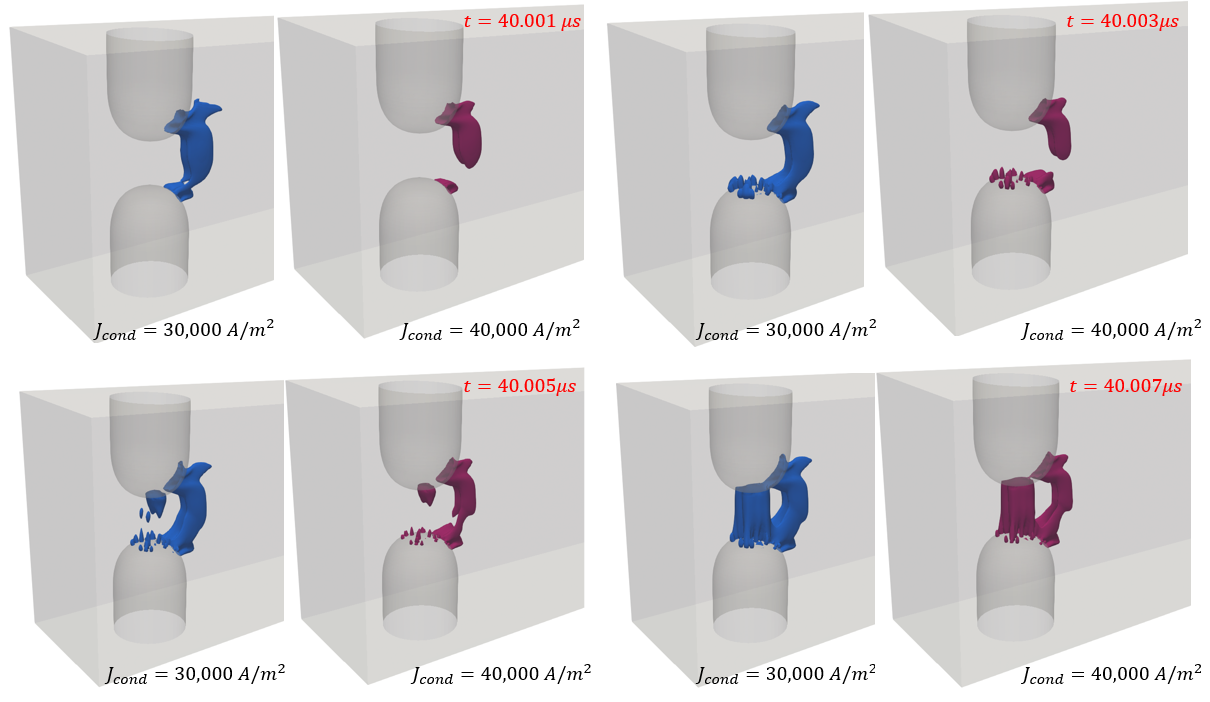}
    \caption{3D isosurface of the current density magnitude ($J_{cond}$) during the third pulse discharge from $t = 40.001\ \mu\text{s}$ to $40.007\ \mu\text{s}$. The blue and maroon structures represent the isosurfaces of $30{,}000\ \text{A/m}^2$ and $40{,}000\ \text{A/m}^2$, respectively.}
    \label{snap-back}
\end{figure}

\begin{figure}[t!]
    \centering
    \includegraphics[width=\linewidth]{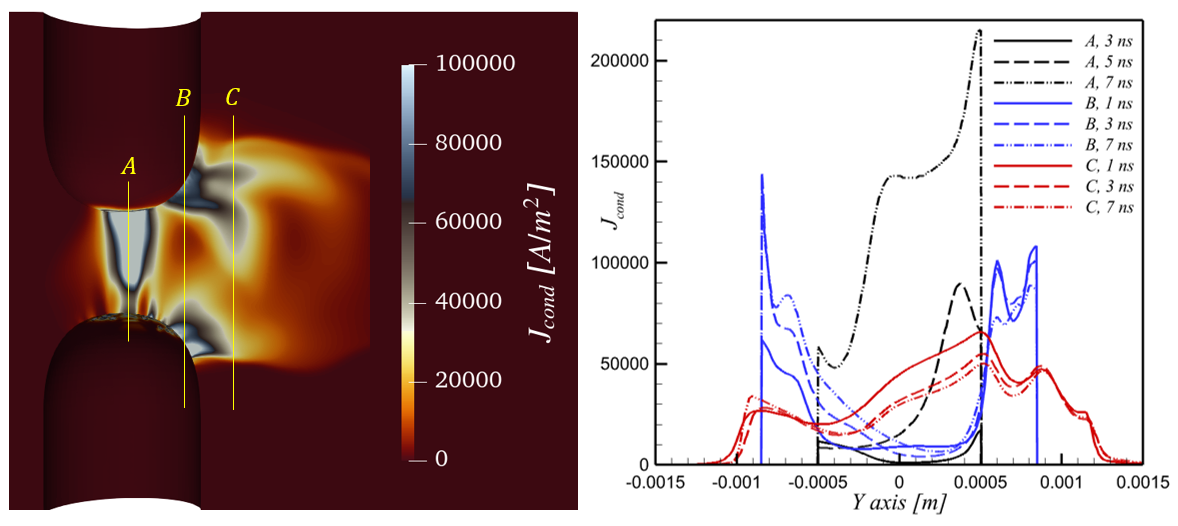}
    \caption{Current density magnitude ($J_{cond}$) at three different positions ($A$, $B$, and $C$) during the third pulse phase (left). Line profiles of $J_{cond}$ as a function of the transverse coordinate $y$, extracted along lines $A, B$, and $C$ at different time instants during the pulse phase ($A$: 3, 5, 7 ns; $B$: 1, 3, 7 ns; $C$: 1, 3, 7 ns) (right).}
    \label{Snap_plot}
\end{figure}

\subsection{Snap-back} 
The electrical current flowing through the channel heats the surrounding gas, causing it to expand and decrease in density, which makes it significantly easier to ionize further. While NRPDs induce localized gas heating, the plasma channel produces high densities of electrons, ions, and excited species for a short period after pulse discharge. Since the electrical resistance of the established plasma channel is significantly lower than that of the surrounding air, subsequent discharge pulses preferentially follow the previous conductive channel rather than forming a new channel. As the previously heated and ionized channel loses conductivity due to electron–ion recombination, the subsequent discharge can shift back toward the shorter, more direct path between the electrodes, resulting in the snap-back phenomenon. 

Figure~\ref{snap-back} illustrates the 3D isosurface of the current density magnitude, $J_{\mathrm{cond}} = |e n_e \mu_e \vec{E}|$, during the third pulse discharge from $t = 40.001\ \mu\text{s}$ to $40.007\ \mu\text{s}$. Two distinct isocontours of the current density magnitude, at $30{,}000\ \text{A/m}^2$ and $40{,}000\ \text{A/m}^2$, are plotted to capture the evolution of the conductive plasma channel. At the onset of discharge ($t = 40.001\ \mu\text{s}$), the $J_{\text{cond}} = 30{,}000\ \text{A/m}^2$ isosurface outlines the conductive channel formed by the second pulse, while the isosurface for $J_{\text{cond}} = 40{,}000\ \text{A/m}^2$ is not fully connected. As the discharge evolves to $t = 40.003\ \mu\text{s}$, a higher current emerges near the cathode surface, driven by intense local impact ionization. At $t = 40.005\ \mu\text{s}$, a positive streamer initiates at the anode and propagates toward the cathode. Finally, at $t = 40.007\ \mu\text{s}$, a new conductive channel forms between the electrodes, and the previous conductive channel is also fully bridged. Consequently, we observe that the current density threshold for snap-back to occur is between $30{,}000\ \text{A/m}^2$ and $40{,}000\ \text{A/m}^2$. Figure~\ref{Snap_plot} plots the evolution of the current density magnitude ($J_{\mathrm{cond}}$) at three different positions ($A$, $B$, and $C$) during the third pulse phase: position~$A$ where the new conductive channel forms directly between the electrodes, position~$B$ where the previous conductive channel resides after being advected downstream by hydrodynamic convection, and position~$C$ where the previous channel terminates. Note that the anode and cathode surfaces at position~$A$ are located at $y = 0.5~\mathrm{mm}$ and $y = -0.5~\mathrm{mm}$, respectively. At position~$C$, the minimum current density within the inter-electrode gap ($-0.5\mathrm{~mm} < y < 0.5\mathrm{~mm}$) decays from $30{,}000\mathrm{~A/m^2}$ to $20{,}000\mathrm{~A/m^2}$ over time, indicating that a subsequent pulse will not conduct through the previous channel. At position~$A$, once a new conductive channel forms, the current density exhibits rapid growth at the anode, increasing from $10{,}000\mathrm{~A/m^2}$ at $t = 3\mathrm{~ns}$ to a peak exceeding $210{,}000\mathrm{~A/m^2}$ at $t = 7\mathrm{~ns}$. At position~$B$ ($-0.9\mathrm{~mm} < y < 0.9\mathrm{~mm}$), the current density near the anode surface decreases as current is redirected into the new conductive channel, whereas near the cathode surface, it rises sharply due to an increased electron density.

\begin{figure}[t!]
    \centering
    \includegraphics[width=\linewidth]{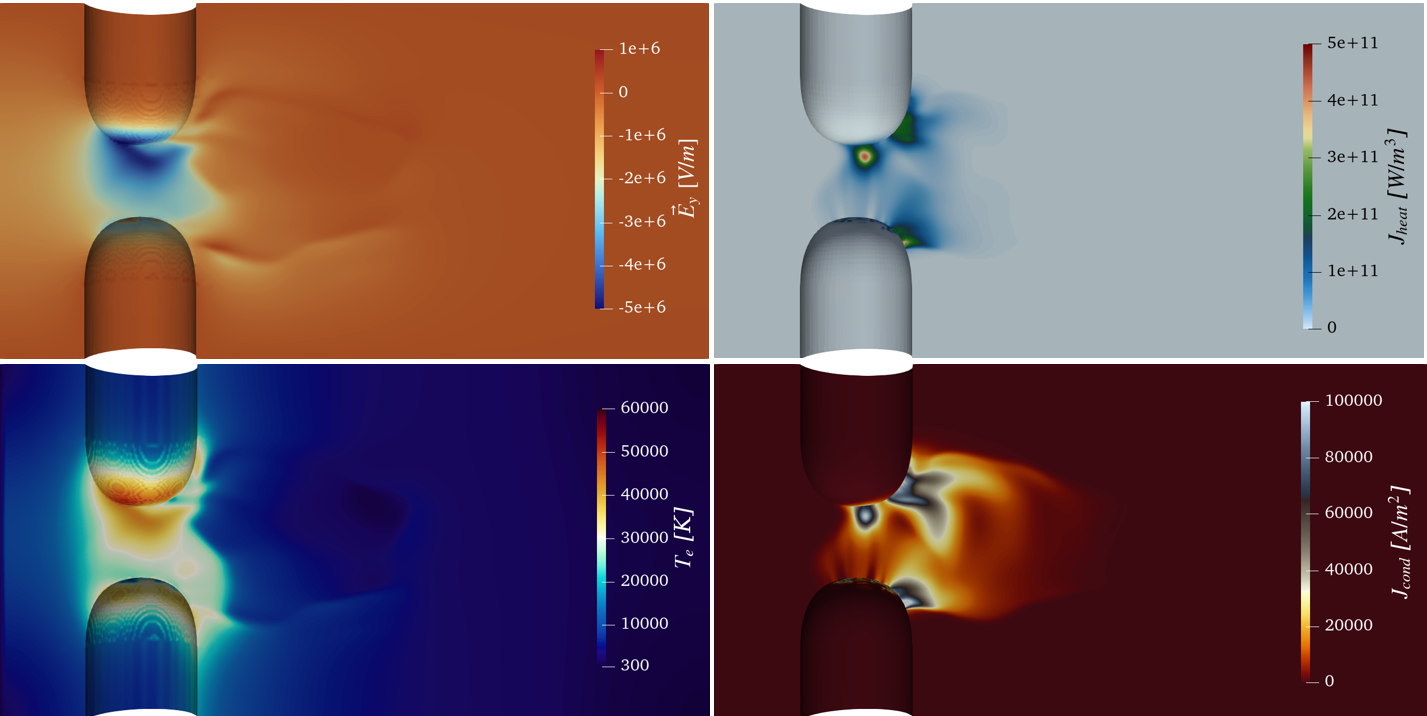}
    \caption{Contours of the vertical electric field component $E_y$ (top left), Joule heating $J_{heat}$ (top right), electron temperature $T_e$ (bottom left), and current density magnitude $J_{\mathrm{cond}}$ (bottom right) at $t = 40.005\ \mu\text{s}$.}
    \label{data_5ns}
\end{figure}

\begin{figure}[h!]
    \centering
    \includegraphics[width=\linewidth]{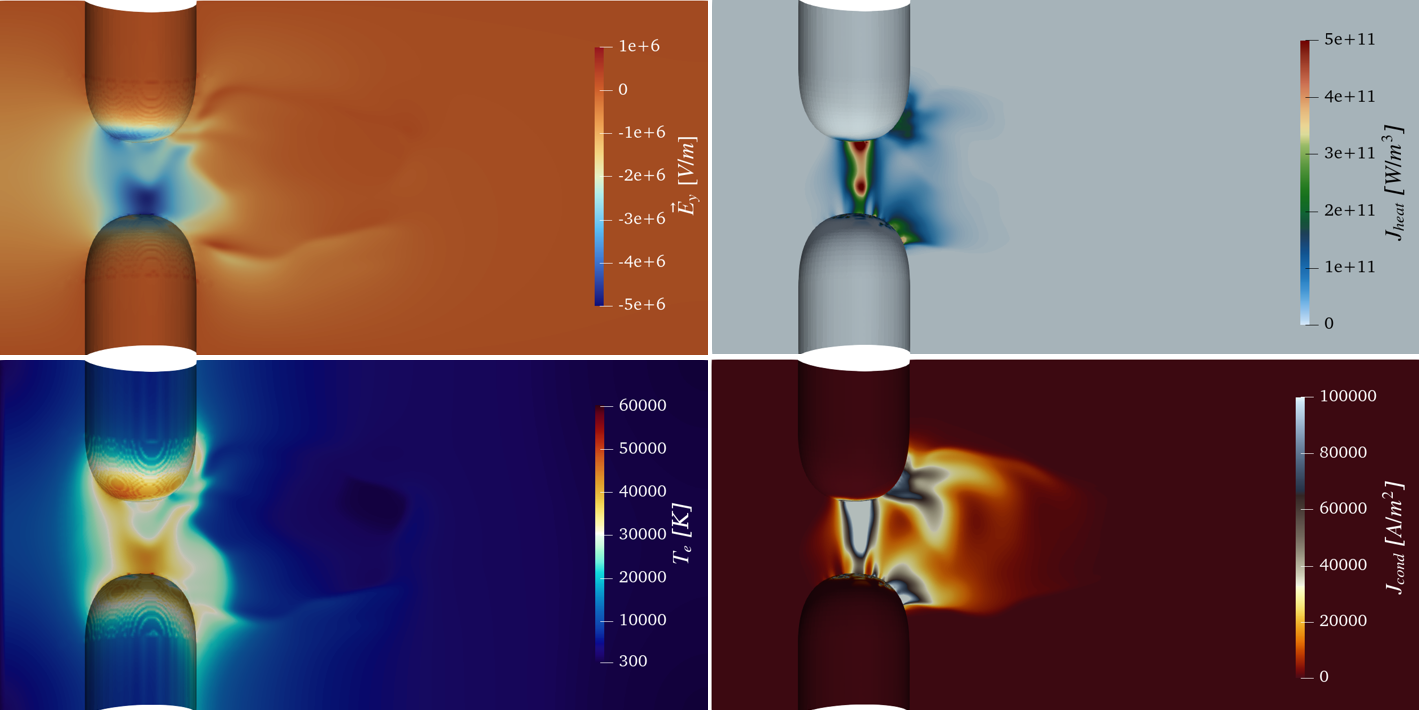}
    \caption{Contours of the vertical electric field component $E_y$ (top left), Joule heating $J_{heat}$ (top right), electron temperature $T_e$ (bottom left), and current density magnitude $J_{\mathrm{cond}}$ (bottom right) at $t = 40.007\ \mu\text{s}$.}
    \label{data_7ns}
\end{figure}

To observe details of development of current, Figure~\ref{data_5ns} showcases contours of the electric field along $y$-axis $\vec{E}_y$, Joule heating $J_{heat}$ ($=-e\vec{\Gamma}_e \cdot \vec{E}$), electron temperature ($T_e$), and current density magnitude $J_{cond}$ ($=|e n_e \mu_e \vec{E}|$) at $40.005\ \mu\text{s}$. As shown, a strong electric field forms between the electrodes during the pulse phase, accelerating electrons and causing intense Joule heating at the anode tip. As a result, the Joule heating increases the local mean electron energy, causing regions of high electron temperature around high electric field regimes. The high electron temperature near the anode contributes to an elevated electron production rate; consequently, a new positive streamer emerges. Figure~\ref{data_7ns} illustrates the same parameters 2 \text{ns} after the time shown in Figure~\ref{data_5ns}. The region of highest electric field moves toward the cathode. Accordingly, the high Joule heating region exhibits a sharp, thin profile directly on the cathode surface, where a locally strong electric field occurs. The electron temperature distribution, which directly affects electron production rates, exhibits broad peaks near the cathode. Consequently, the intense current generated by the electron avalanche near the cathode establishes a new path of least resistance directly between the electrodes, triggering the snap-back.

\section{Conclusion} \label{s_conclusion} 
This work develops a detailed plasma-assisted combustion simulation framework based on drift-diffusion-convection modeling, \textit{PlasmaCombustionFoamLM} (or \textit{PCFoamLM}), to investigate the plasma morphology under nanosecond repetitive pulsed discharges (NRPDs) in air cross‑flows. By coupling drift–diffusion–convection transport, electrostatic potential, photoionization, detailed plasma chemistry, a mean electron energy formulation, and ambipolar diffusion within OpenFOAM, the solver captures the multi-scale physics of streamer initiation, propagation, and afterglow across both pulse and gap phases. An adaptive time-stepping strategy is implemented to resolve the highly disparate timescales inherent to the coupled electrostatic, drift-diffusion-convection, and chemical kinetic processes, enabling stable integration across multiple pulse and gap phases. Verification and validation against three benchmark cases are performed, confirming the accuracy and robustness of the solver for plasma-assisted combustion (PAC). 

Using the present simulation framework, we investigate plasma morphology in a 3D channel with a pin-to-pin electrode configuration under two representative cross-flow velocities (40 and 100 m/s) at a fixed pulse repetition frequency of 50 kHz. The key findings of this study are summarized as follows:
\begin{itemize} 
    \item At a lower cross-flow velocity of $40\text{ m/s}$, successive pulses visibly increase the electron density, which persists in the pre-existing channel near the anode. This persistent electron density promotes continued energy deposition into the conductive channel, driving substantial thermal accumulation and establishing a continuous high-temperature channel downstream that stabilizes extended reaction zones.  
    
    \item At a higher cross-flow velocity of $100\text{ m/s}$, however, due to the significant decrease in electron and ion densities between pulses (i.e., a decrease in plasma retention), thermal accumulation downstream due to energy deposition from the previously formed conductive channels becomes weaker that the peak temperature is lower than that at the lower cross-flow velocity. 

    \item The current density threshold for snap-back—where the previously heated and ionized channel loses conductivity due to electron–ion recombination, causing the subsequent discharge to shift back toward the shorter, more direct path between the electrodes—is between $30{,}000\ \text{A/m}^2$ and $40{,}000\ \text{A/m}^2$.

    \item The physical mechanism of snap-back begins when a strong electric field forms between the electrodes, accelerating electrons and causing intense Joule heating at the anode tip. This elevates the local electron temperature, which boosts the electron production rate so that more intense current between the electrodes is generated, which establishes a new path of least resistance directly between the electrodes.
\end{itemize}
 
Beyond identifying the dominant physics governing plasma morphology and snap-back, this study highlights the importance of accurately capturing streamer propagation, photoionization, afterglow, ambipolar diffusion, and residual-channel evolution when modeling NRPDs. These insights provide a robust, physics‑based foundation for optimizing NRPD actuators in demanding applications such as plasma‑assisted combustion and ignition enhancement. The modeling framework developed here is particularly well‑suited for extension to reactive fuel–air environments. However, detailed plasma and combustion chemistry in three‑dimensional simulations necessitates careful mechanism reduction to manage high computational costs. Future work will extend this framework to fully three‑dimensional simulations of fuel–air mixtures in channel flows to examine flame‑kernel evolution and the coupling between plasma energy deposition and early‑stage ignition dynamics. In addition, the solver will be applied to plasma‑enhanced flame holding in scramjet, where the ability of NRPDs to sustain stable combustion in high‑speed, cross‑flow environments is critical for improving engine performance, operability, and ignition reliability.

\ack{S. Yang gratefully acknowledges the support from the Office of Naval Research (ONR) Grant under Award Number N00014-26-1-2321 under the supervision of project monitors Dr. Eric Marineau and Dr. Jonathan Sosa. The information, data, or work presented herein was also funded in part by the National Science Foundation (NSF) under Award Number CBET 2002635. The views and opinions of authors expressed herein do not necessarily state or reflect those of the United States Government or any agency thereof. G. Shim gratefully acknowledges the financial support of the Abplanalp-Sanders Memorial Fellowship from the University of Minnesota (UMN). 
The authors also acknowledge the Minnesota Supercomputing Institute (MSI) for its computational resources. The authors used Google Gemini during the preparation of this manuscript for grammar checking, and improving the clarity of the text. The authors reviewed and critically assessed all AI-assisted outputs and take full responsibility for the content and accuracy of the manuscript.}

\section*{Data availability statement}
The data that support the findings of this study are available from the corresponding author upon reasonable request.


\appendix 
\phantomsection
\section{Appendix: Solver Verification \& Validation} \label{appendix}
In this section, we assess the accuracy and reliability of the drift–diffusion–convection plasma–fluid solver through verification and validation cases. Three representative problems are selected to evaluate the numerical implementation. The results obtained from our solver are compared against published experimental measurements and benchmark numerical data, demonstrating the solver’s capability to reproduce key plasma‑physics behavior under atmospheric conditions.

\begin{figure}[h]
    \centering

    \begin{subfigure}{0.9\linewidth}
        \centering
        \includegraphics[width=\textwidth,trim=3 3 3 3,clip]{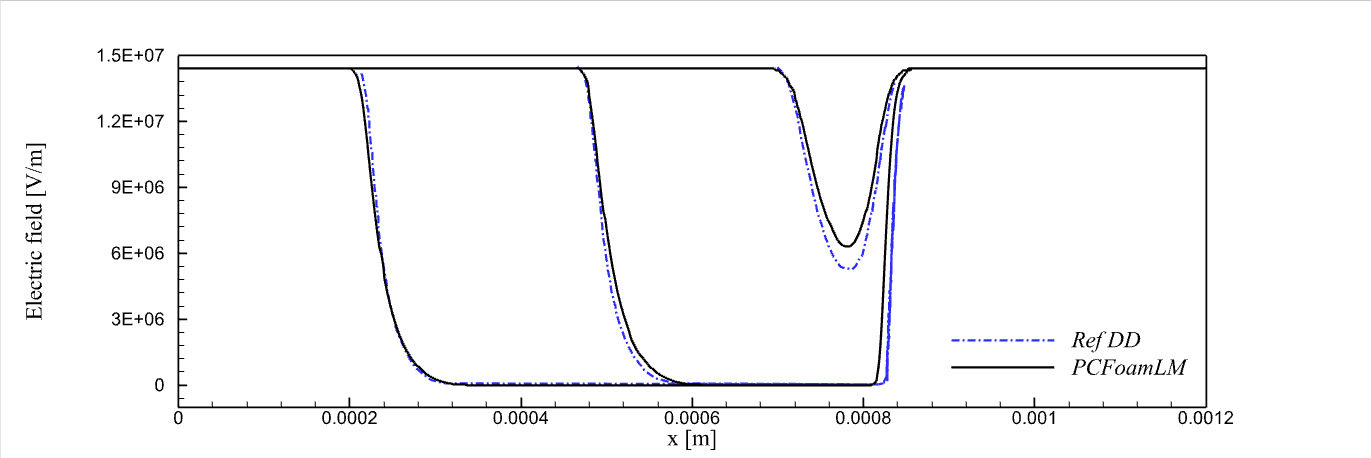}
        \caption{}
    \end{subfigure}
    \begin{subfigure}{0.9\linewidth}
        \centering
        \includegraphics[width=\textwidth,trim=3 3 3 3,clip]{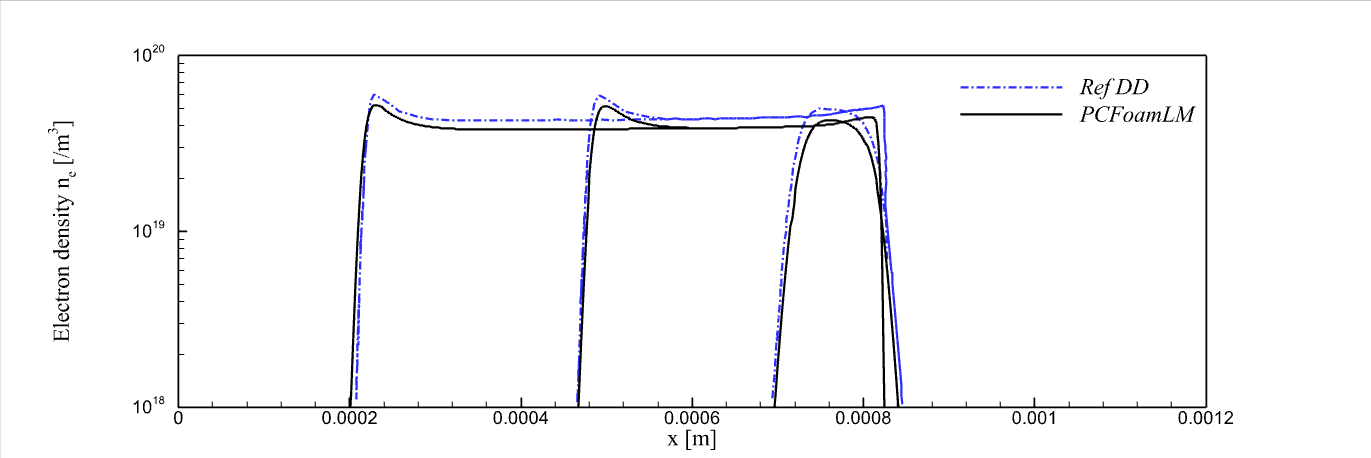}
        \caption{}
    \end{subfigure}
    \begin{subfigure}{0.9\linewidth}
        \centering
        \includegraphics[width=\textwidth,trim=3 3 3 3,clip]{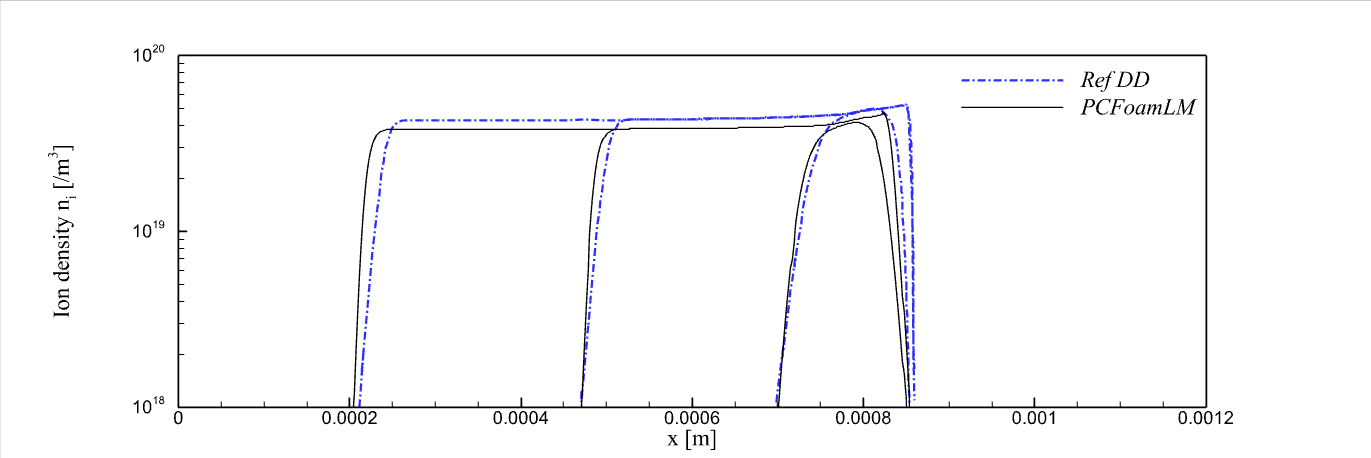}
        \caption{}
    \end{subfigure}

    \caption{Comparison of the electric field, electron number density, and ion number density between the drift–diffusion model of Markosyan~\textit{et al.} (\textit{Ref DD}) and the current solver (\textit{PCFoamLM}) at three different times ($t=0.07$, 0.35, and $0.7\;\mathrm{ns}$). (a) Electric field, (b) electron number density, and (c) ion number density.}
    \label{negativeStreamer}
\end{figure}

\subsection{A negative (anode-directed) streamer in nitrogen} 
A negative (anode-directed) one-dimensional streamer ionization front in nitrogen is compared against results from the first-order fluid model of Markosyan~\textit{et al.}~\cite{markosyan2013high}. The simulation domain consists of a one-dimensional inter-electrode gap of length $1.2\,\mathrm{mm}$, discretized using $1025$ uniformly spaced grid points. The background gas is nitrogen at $298\,\mathrm{K}$ and $1\,\mathrm{atm}$, with a reduced electric field ($E/N$) of $590\,\mathrm{Td}$ applied across the gap. A fixed time step of $0.05\,\mathrm{ps}$ is used throughout the simulation. The anode is held at $17.3\,\mathrm{kV}$, while the cathode is grounded. Boundary conditions consist of a homogeneous Neumann condition at the anode and a Dirichlet condition at the cathode for the electron number density. To initiate streamer development, a Gaussian seed of electrons and ions is placed near the cathode, defined as:
\begin{equation}
n_{e/i}(x)\big|_{t=0} = n_{e0/i0} \exp\left( \frac{-(x - x_0)^2}{\sigma^2} \right),
\label{eq:gaussian_seed}
\end{equation}
where $n_{e0/i0} = 2 \times 10^{18}\,\mathrm{m}^{-3}$, $\sigma = 0.029\,\mathrm{mm}$, and $x_0 = 0.8\,\mathrm{mm}$. As the local field approximation (LFA) is employed here, the ionization rate coefficient, electron mobility, and diffusion coefficients are tabulated as functions of the local reduced electric field~\cite{taneja20241d}. All electron transport and reaction properties depend solely on the instantaneous value of $E/N$. The reaction rate coefficient $k$ for electron-impact ionization (\ce{N2} + \ce{e-} $\rightarrow{}$ \ce{N2+} + 2\ce{e-}) is given by
\begin{align}
\log_{10}k &= -218.244 + (316.770)\xi + (-203.767)\xi^2 + (66.912)\xi^3 + (-11.079)\xi^4 + (0.735)\xi^5, 
\end{align} 
where $\xi = \text{log}_{10}(E/N)$. The electron mobility is obtained from a piecewise fit:
\begin{align}
\mu_e = 
\begin{cases} 
-0.1811 & \text{for } \xi < 0.3 \\
-\frac{1.0 \times 10^{24}}{N}
\left(
\begin{aligned}
&(1063.323) + (-1197.375)\xi \\
& + (691.236)\xi^2 + (-164.0486)\xi^3 \\
& + (12.995)\xi^4 + (-16.851)\log(\xi) \\
&+ (-47.511)\sin(\xi) + (226.179)\cos(\xi) \\
& - (1322.495)\exp(-\xi)
\end{aligned}
\right) & \text{for } 0.3 < \xi < 3.0 \\
-0.0188 & \text{for } \xi > 3.0
\end{cases} 
\end{align} 
Note that the minus sign appears because the drift–diffusion formulation adopts a positive‑sign convention, causing the electron mobility term to carry a minus sign to reflect the opposite charge polarity. The electron diffusion coefficient is obtained using Einstein’s relation, $D_e = \mu_e k_B T_e/e$. Since the total simulation time is 1 ns, we assume that the heavy species, \ce{N2} and \ce{N2+}, remain frozen in contrast to the short drift--diffusion timescale of electrons. Therefore, the drift--diffusion terms of the heavy species are negligible. Here, a first-order Euler scheme is employed for temporal discretization. The gradient terms are discretized using the Gauss linear scheme, while the divergence term for electron density is discretized using the first-order upwind scheme. Diffusion terms are evaluated using the Gauss linear corrected scheme.

Figure~\ref{negativeStreamer} showcases the electric fields and the electron and ion number densities at three different times ($t=0.07$, 0.35, and $0.7\;\mathrm{ns}$) obtained using \textit{PCFoamLM} solver, compared with the drift–diffusion model of Markosyan~\textit{et al.}~\cite{markosyan2013high}. A strong background electric field ($145\;\mathrm{kV cm^{-1}}$) initiates electron-impact ionization by accelerating free electrons to energies above the ionization threshold over sub-micron distances, thereby triggering an avalanche and rapidly increasing both electron and ion densities. The electrons then drift toward the anode while the heavier ions remain stationary over the electron drift time scale. This imbalance generates a space-charge field that progressively distorts the electric field, thereby promoting the onset of ionization-front propagation.

\subsection{Streamer propagation with photoionization} \label{photo_appendix}
In order to observe the effect of photoionization, we neglect ion transport over the short drift–diffusion timescale of electrons, as done in Bagheri~\textit{et al.}~\cite{bagheri2018comparison}. The electron and positive ion densities evolve over time and are coupled to the electrostatic potential through Poisson’s equation:
\begin{align}
 & \frac{\partial n_e}{dt} = \nabla \cdot \left(n_e \mu_e \vec{E} + D_e \nabla n_e \right) + \bar{\alpha} \mu_e |\vec{E}| n_e + S_{ph}, \\
 &\frac{\partial n_i}{dt} = \bar{\alpha} \mu_e |\vec{E}| n_e + S_{ph}, \\
 & \nabla^2 \phi_e + \frac{e}{\epsilon_0} \sum_{i=1}^{N_{sp}} Z_i n_i = 0, \quad \vec{E} = - \nabla \phi_e
\end{align}  
where $\bar{\alpha}$ denotes the effective ionization coefficient. According to the Bagheri~\textit{et al.}~\cite{bagheri2018comparison}, the transport and ionization coefficients are given by
\begin{align}
\begin{split}
 \mu_e &= 2.3987 |\vec{E}|^{-0.26}, \\
 D_e &= \num{4.3628e-3} |\vec{E}|^{0.22}, \\
 \alpha &= \left(\num{1.1944e6} + \frac{4.3666 \times 10^{26}}{|\vec{E}|^3}\right) \text{exp}\left(-\frac{\num{2.73e7}}{ |\vec{E}|}\right) \\
 \eta &= 340.75 \\
 \bar{\alpha} &= \alpha - \eta       \label{transport_photo}
\end{split}
\end{align} 

\begin{figure}
    \begin{subfigure}{0.5\linewidth}
        \centering
        \includegraphics[width=\textwidth,trim=3 3 3 3,clip]{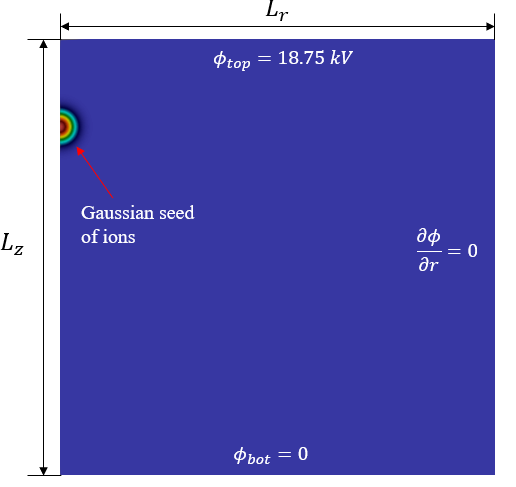}
        \caption{}
        \label{Domain_photo}
    \end{subfigure}
    \begin{subfigure}{0.354\linewidth}
        \centering
        \includegraphics[width=\textwidth,trim=3 3 3 3,clip]{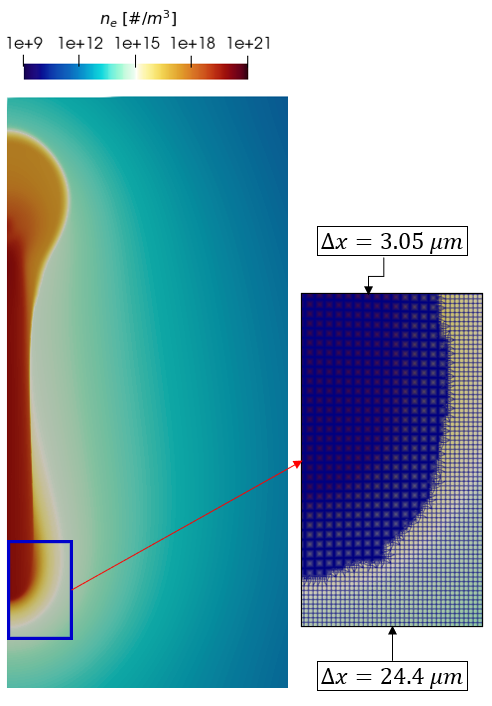}
        \caption{}
        \label{AMR}
    \end{subfigure}
    \caption{(a) Geometry and boundary conditions for the electrostatic potential in the 2D axisymmetric domain. A Gaussian ion seed is placed at the upper-left region, with $\phi_{top}=18.75\,\mathrm{kV}$, $\phi_{bot}=0$, and zero-gradient on the left and right boundaries. (b) Electron density $n_e$, including a zoomed-in view using adaptive mesh refinement (AMR). The AMR spans cell sizes from $\Delta x = 24.4~\mathrm{\mu m}$ down to $\Delta x = 3.05~\mathrm{\mu m}$ over three refinement levels.}
\end{figure}

Figure~\ref{Domain_photo} illustrates a 2D axisymmetric domain with dimensions $L_r = L_z = 1.25\;\mathrm{cm}$. A potential of $\phi_{top} = 18.75\;\mathrm{kV}$ is applied at the top boundary, while the bottom boundary is grounded ($\phi_{bot} = 0$). 
For the left and right boundaries, a homogeneous Neumann condition is imposed on the potential ($\partial_r \phi = 0$). Immobile positive ions are seeded with a Gaussian distribution as follows:
\begin{align}
 n_{i}(r,z)\big|_{t=0}&= N_0 \; \text{exp}\left[-\frac{r^2+\left(z-z_0\right)^2}{\sigma^2} \right].
\end{align} 
where $N_0=\num{5e18}\;\mathrm{m^{-3}}, \sigma = \num{4e-4}\;\mathrm{m}$, and $z_0=\num{1e-2}\;\mathrm{m}$. 

OpenFOAM provides adaptive mesh refinement (AMR) through its native dynamicRefineFvMesh framework, which enables on‑the‑fly refinement and coarsening of 3D hexahedral meshes based on user‑defined criteria. However, the native AMR capability is restricted to structured hexahedral meshes and does not support true 2D refinement. In this work, we employ an extended 2D AMR framework provided by an external library~\cite{sun2023detonationfoam}. Figure~\ref{AMR} shows the electron density field, along with a zoomed-view illustrating the locally adaptive mesh refinement (AMR). The computational domain initially contains 262{,}144 ($512 \times 512$) cells, with $\Delta x = 24.4\;\mathrm{\mu m}$. To resolve the space‑charge layers around the streamer head and the regions with high electric field, three levels of AMR are employed, refining the mesh down to $\Delta x = 3.05\;\mathrm{\mu m}$. To observe the effects of photoionization on positive streamers in dry air at 1 bar and 300 K, two different background levels are tested, with and without photoionization included. Here, a first-order Euler scheme is employed for temporal discretization. The gradient terms are discretized using the Gauss linear scheme, while the divergence term for electron density is discretized using the first-order upwind scheme. Here, a first-order Euler scheme is employed for temporal discretization. The gradient terms are discretized using the Gauss linear scheme, while the divergence term for electron density is discretized using the first-order upwind scheme. Diffusion terms are evaluated using the Gauss linear corrected scheme. The Poisson's equation for the electrostatic potential and the Helmholtz equations for photoionization are solved using the GAMG (Geometric-Algebraic Multi-Grid) solver, which is well suited for solving large sparse linear systems arising from elliptic equations.

\begin{table}[h!]
\centering
\caption{Parameters for the three-exponential fit of the absorption function taken from Bourdon~\textit{et al.}~\cite{bourdon2007efficient}.}
\label{tab:threeExp}
\begin{tabular}{|c|c|c|c|}
\hline
$j \rightarrow$ & $j = 1$ & $j = 2$ & $j = 3$ \\
\hline
$A_j~(\text{cm}^{-2} \text{Torr}^{-2})$ & $1.986 \times 10^{-4}$ & $5.1 \times 10^{-3}$ & $4.886 \times 10^{-1}$ \\
$\lambda_j~(\text{cm}^{-1} \text{Torr}^{-1})$ & $5.53 \times 10^{-2}$ & $1.46 \times 10^{-1}$ & $8.9 \times 10^{-1}$ \\
\hline
\end{tabular}
\end{table}

\subsubsection{Case 1. High pre-ionization without photoionization} 
\begin{figure}
    \centering
    \begin{subfigure}{0.415\linewidth}
        \centering
        \includegraphics[width=\textwidth,trim=3 3 3 3,clip]{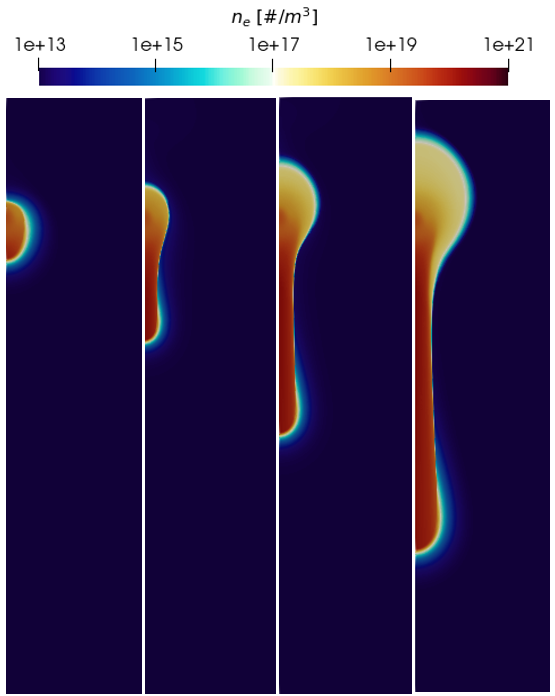}
        \caption{}
    \end{subfigure}
    \begin{subfigure}{0.4\linewidth}
        \centering
        \includegraphics[width=\textwidth,trim=3 3 3 3,clip]{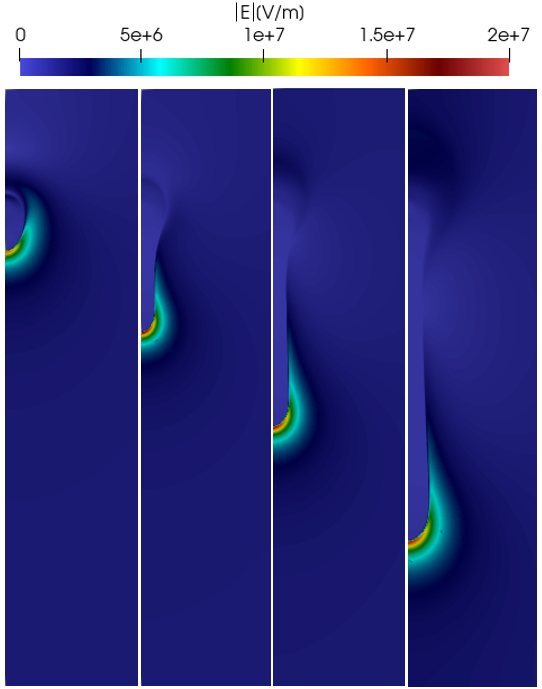}
        \caption{}
    \end{subfigure}
    \caption{Evolution of the positive streamer at four times (2, 6, 10, and 14 ns) under high pre‑ionization ($n_e=n_i=10^{13}\mathrm{m^{-3}}$) without photoionization: (a) electron density $n_e$, (b) electric field magnitude $|\vec{E}|$.}
    \label{wo_photo}
\end{figure}

\begin{figure}[t!]
    \centering
    \begin{subfigure}{0.49\linewidth}
        \centering
        \includegraphics[width=\textwidth,trim=15 15 15 15,clip]{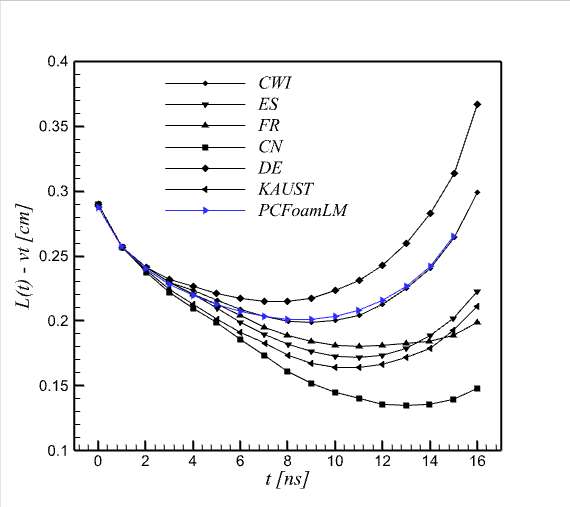}
        \caption{}
    \end{subfigure}
    \begin{subfigure}{0.49\linewidth}
        \centering
        \includegraphics[width=\textwidth,trim=15 15 15 15,clip]{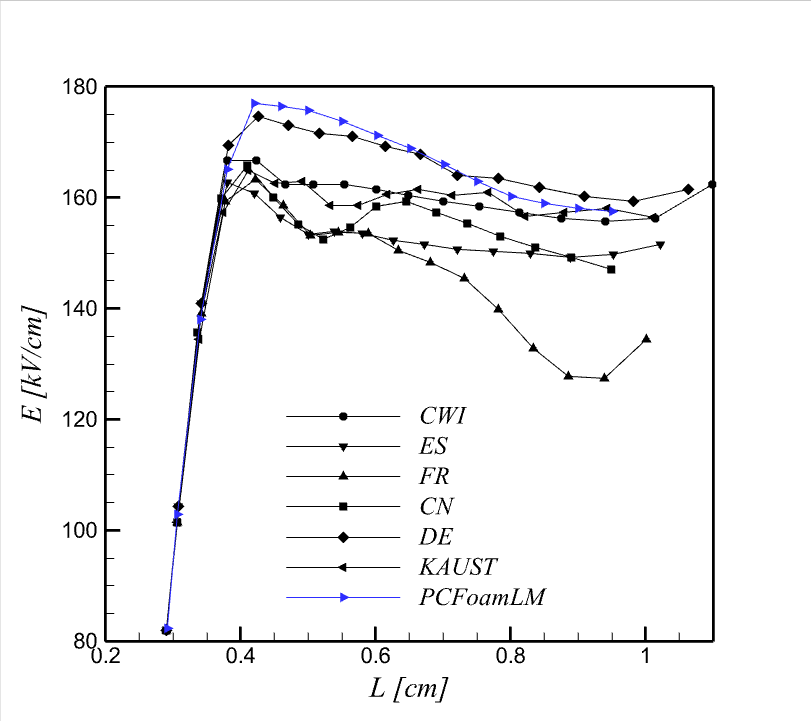}
        \caption{}
    \end{subfigure}
    \caption{Comparison of the streamer length $L(t)$ and electric field magnitude $|\vec{E}|$ under high pre‑ionization ($n_e=n_i=10^{13}\mathrm{m^{-3}}$) without photoionization. \textit{PCFoamLM} (blue) is compared with results reported by other groups. (a) streamer length $L(t)-\nu t$, with $\nu = 0.05 \; \mathrm{cm \; ns^{-1}}$, (b) electric field magnitude $|E|$ as a function of streamer length $L$.}
    \label{wo_nE}
\end{figure}

The first case employs a homogeneous background ionization level of $n_e = n_i = \num{1e13}\;\mathrm{m^{-3}}$ without a photoionization source. A high background electron density reduces gradients in both electron and ion concentrations and lowers the electric field at the streamer tip, making the simulations easier to perform from a computational standpoint. A refinement criterion is applied: the mesh is refined when $\alpha(|\vec{E}|) \Delta r > 0.8$ and de-refine when $\alpha(|\vec{E}|) \Delta r < 0.8$~\cite{teunissen2017simulating}.  The simulations employ a fixed time step of $\Delta t= 3\;\mathrm{ps}$ use up to approximately 730{,}000 computational cells. Figure~\ref{wo_photo} illustrates the evolution of the electron density and electric field at four different times: 2, 6, 10, and 14 ns. The high voltage applied at the top (anode) generates a strong electric field that drives electrons downward toward the bottom (cathode). Consequently, electrons move much faster than ions, drift and diffusion create charge separation at the streamer front, which in turn produces the high electric field at the streamer head. Note that the electric field behind the streamer head is nearly zero, as an equal amount of ions is produced according to the ion conservation equation. Figure~\ref{wo_nE} represents the streamer length $L(t)$ and magnitude of electric field $|E|$ as a function of $L$, compared with results reported by other groups. Here, the streamer length is defined as
\begin{align}
 & L(t) = 0.125 - z_{max}(t) \; [\mathrm{cm}]
\end{align} 
where $z_{max}$ is the axial location of the maximum electric field magnitude. The streamer length predicted by \textit{PCFoamLM} is identical to that obtained with the Afivo‑streamer code from \textit{CWI}~\cite{teunissen2017simulating}, an open‑source plasma‑fluid solver for streamer‑discharge simulations. The electric field magnitude increases sharply to 177 kV/cm due to strong charge separation at the streamer front, and then gradually decreases by about 160 kV/cm when $L=0.9\;\mathrm{cm}$.

\subsubsection{Case 2. Low pre-ionization with photoionization}
\begin{figure}[t!]
    \centering
    \begin{subfigure}{0.305\linewidth}
        \centering
        \includegraphics[width=\textwidth,trim=3 3 3 3,clip]{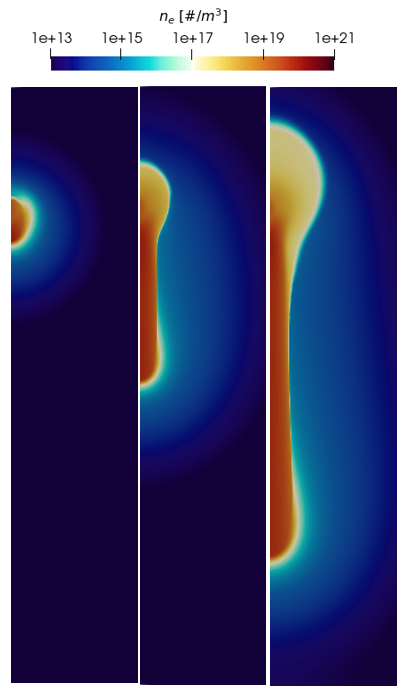}
        \caption{}
    \end{subfigure}
    \begin{subfigure}{0.32\linewidth}
        \centering
        \includegraphics[width=\textwidth,trim=3 3 3 3,clip]{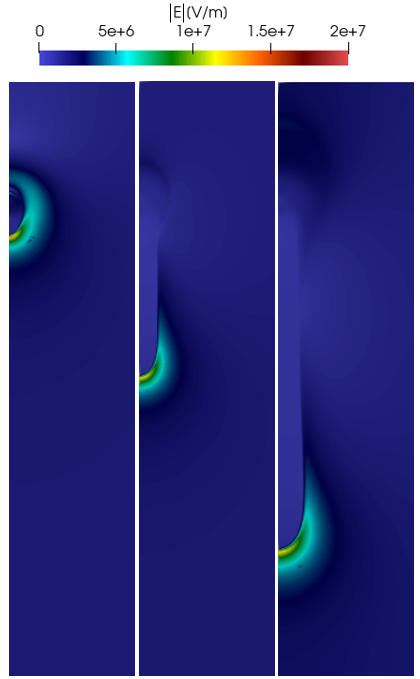}
        \caption{}
    \end{subfigure}
    \begin{subfigure}{0.32\linewidth}
        \centering
        \includegraphics[width=\textwidth,trim=3 3 3 3,clip]{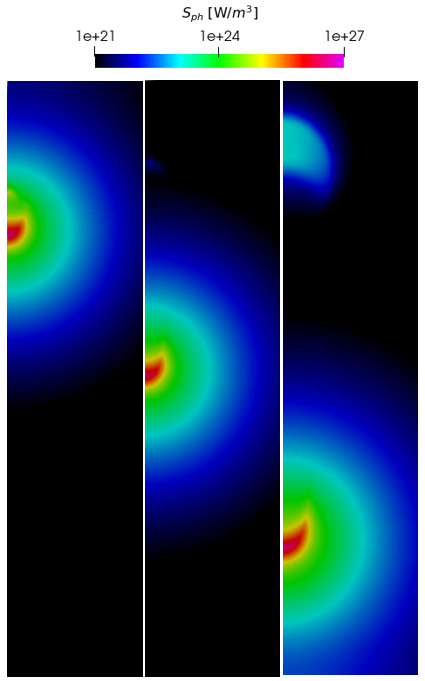}
        \caption{}
    \end{subfigure}
    \caption{Evolution of the positive streamer at three times (2, 8, and 14 ns) under low pre‑ionization ($n_e=n_i=10^{9}\mathrm{m^{-3}}$) with photoionization: (a) electron density $n_e$, (b) electric field magnitude $|\vec{E}|$, (c) photoionization source $S_{ph}$ (log-scale).}
    \label{w_photo}
\end{figure}

\begin{figure}[t!]
    \centering
    \begin{subfigure}{0.49\linewidth}
        \centering
        \includegraphics[width=\textwidth,trim=15 15 15 15,clip]{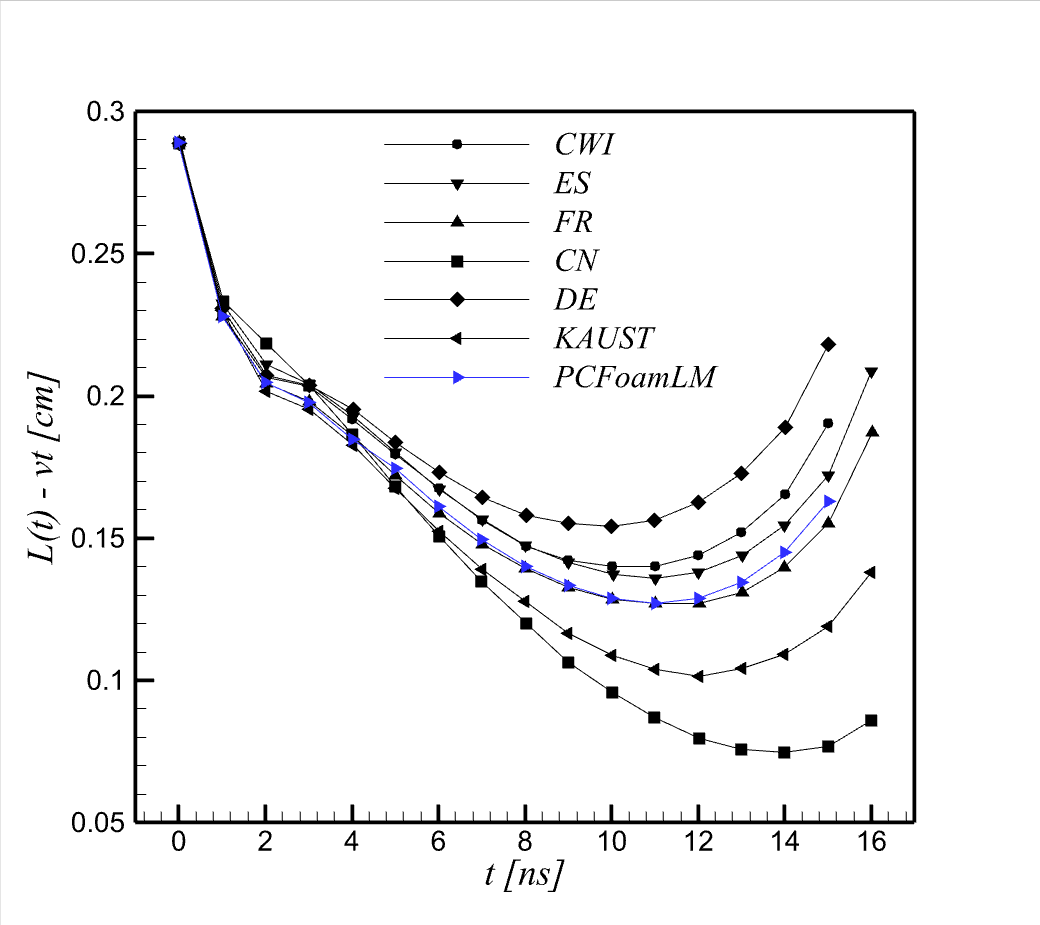}
        \caption{}
    \end{subfigure}
    \begin{subfigure}{0.49\linewidth}
        \centering
        \includegraphics[width=\textwidth,trim=15 15 15 15,clip]{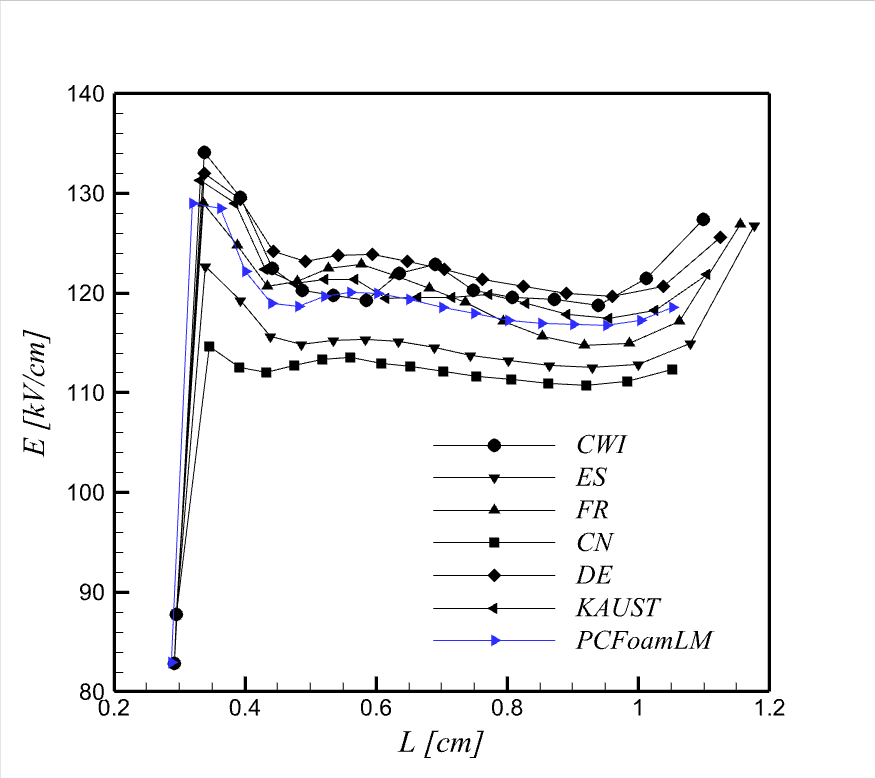}
        \caption{}
    \end{subfigure}
    \caption{Comparison of the streamer length $L(t)$ and electric field magnitude $|\vec{E}|$ under low pre‑ionization ($n_e=n_i=10^{9}\mathrm{m^{-3}}$) with photoionization. \textit{PCFoamLM} (blue) is compared with results reported by other groups. (a) streamer length $L(t)-\nu t$, with $\nu = 0.06 \; \mathrm{cm \; ns^{-1}}$, (b) electric field magnitude $|E|$ as a function of streamer length $L$.}
    \label{w_nE}
\end{figure}

HThe second case employs a homogeneous background ionization level of $n_e = n_i = \num{1e9}\;\mathrm{m^{-3}}$, together with a photoionization source, in dry air at 1 bar and 300 K. A refinement criterion with photoionization is applied: the mesh is refined when $\alpha(|\vec{1.2E}|) \Delta r > 1.0$ and de‑refined when $\alpha(|\vec{1.2E}|) \Delta r < 1.0$~\cite{teunissen2017simulating}. For the photoionization source term $S_{\text{ph}}^j(\vec{r})$, we employ $p_q=40\;\mathrm{mbar}$, $p=750\;\mathrm{Torr}=1\;\mathrm{bar}$, $\zeta(\nu_u/\nu_i) = 0.075$, $S_i(\vec{r})= \alpha \mu_e |\vec{E}| n_e$, and $p_{\ce{O2}}=150\;\mathrm{Torr}$~\cite{bagheri2018comparison}. The three‑exponential fit of the absorption function is given in Table~\ref{tab:threeExp}. Figure~\ref{w_photo} illustrates the evolution of the electron density, electric field, and photoionization source at three different times: 2, 8, and 14 ns. Since UV photons generate seed electrons ahead of the ionization front, they partially neutralize the strong space-charge layer that normally forms at the streamer head, resulting in a lower electric-field peak than in the case without photoionization. Figure~\ref{w_nE} shows the streamer length $L(t)$ and magnitude of electric field $|E|$ as a function of $L$ with photoionization, compared with results reported by other groups. The results exhibit good agreement with previously reported data in terms of both the streamer length and the electric‑field magnitude. 

\begin{figure}[b!]
    \centering
    \includegraphics[width=\linewidth]{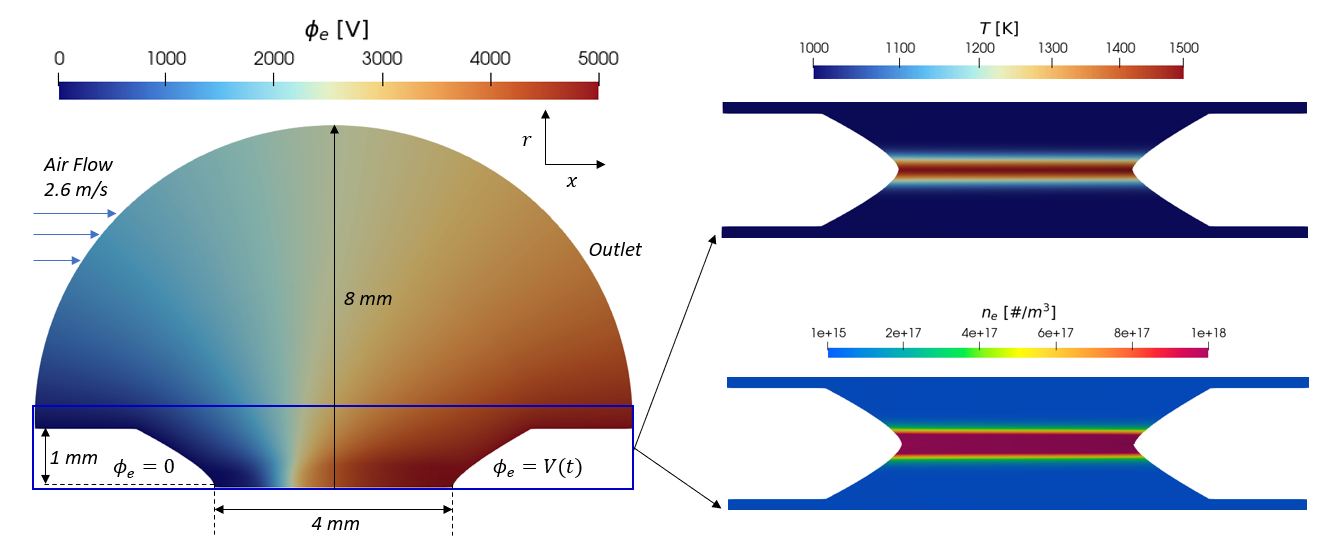}
    \caption{Geometry and initial conditions for the 2D axisymmetric domain between two electrodes. Radial Gaussian profiles for the temperature and electron density across the gap are imposed to represent the effects of multi‑pulse excitation from preceding pulses.}
    \label{Rustholtz_domain}
\end{figure}

\begin{figure}
    \centering
    \includegraphics[width=\linewidth]{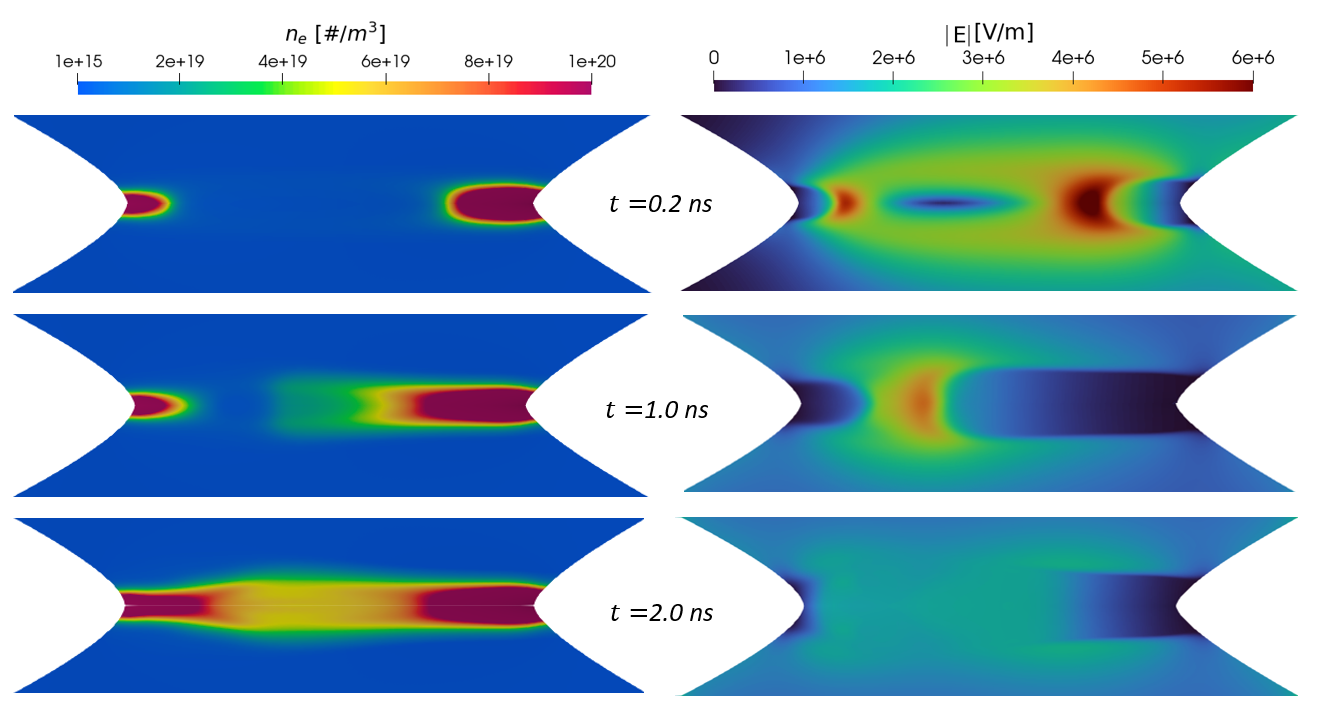}
    \caption{Temporal evolution of the electron density $n_e$ and electric‑field magnitude $|\vec{E}|$ within the constricted channel at $t = 0.2, 1.0$, and 2.0 ns. The early high‑field region (highlighted at 0.2 ns) initiates localized ionization, followed by streamer propagation and the formation of a conductive plasma channel that smooths the electric field and broadens the electron density profile.}
    \label{Rustholtz_streamer}
\end{figure}

\begin{figure}
    \centering
    \begin{subfigure}{0.49\linewidth}
        \centering
        \includegraphics[width=\textwidth,trim=15 15 15 15,clip]{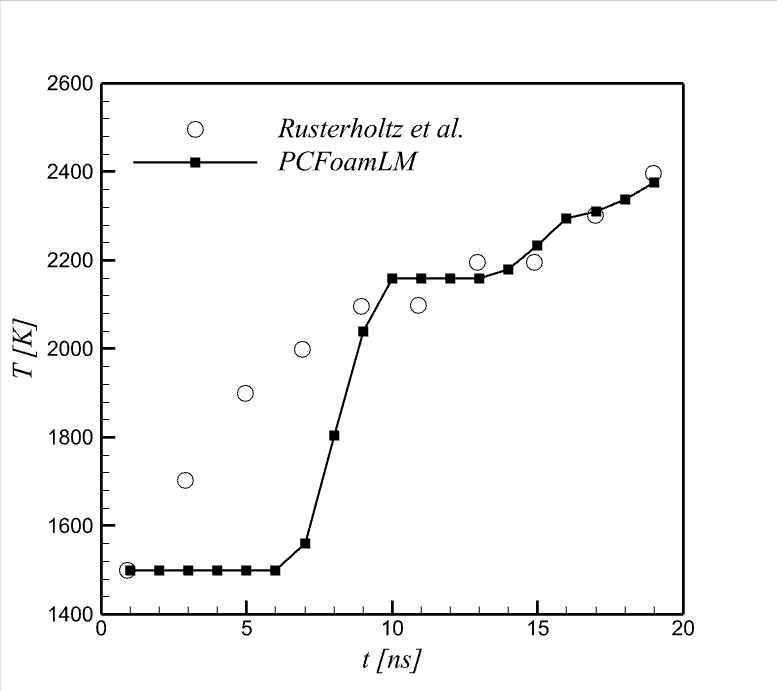}
        \caption{}
    \end{subfigure}
    \begin{subfigure}{0.49\linewidth}
        \centering
        \includegraphics[width=\textwidth,trim=15 15 15 15,clip]{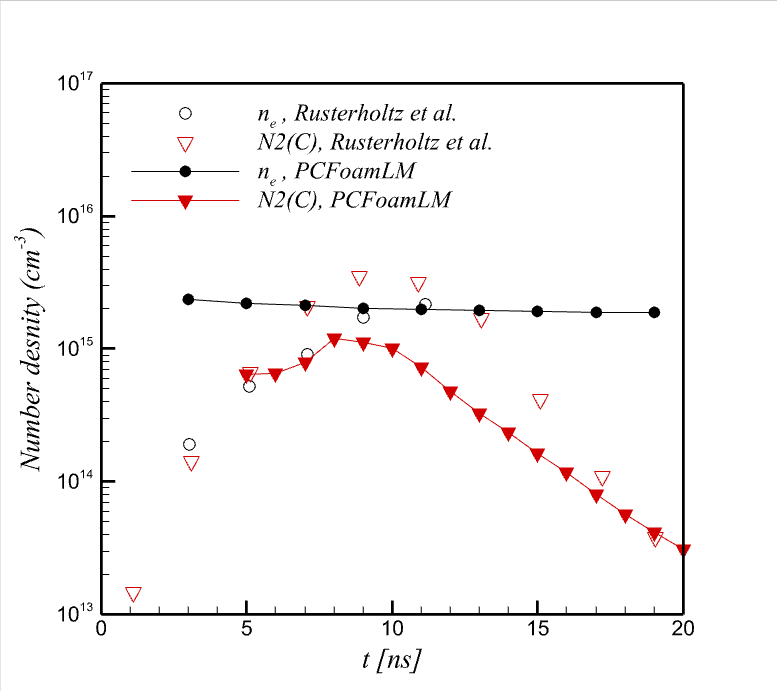}
        \caption{}
    \end{subfigure}
    \caption{Comparison of numerical results from \textit{PCFoamLM} with experimental data from Rusterholtz~\textit{et al.}~\cite{rusterholtz2013ultrafast} in atmospheric air. (a) Maximum gas temperature $T$. (b) Maximum electron number density ($n_e$) and the number density of the nitrogen excited state $N_2(C)$, both measured in $cm^{-3}$.}
    \label{Rustholtz_validation}
\end{figure}

\subsection{Nanosecond
repetitively pulsed discharges in air} 
In atmospheric air, Rusterholtz~\textit{et al.}~\cite{rusterholtz2013ultrafast} conducted a detailed experimental investigation of ultrafast heating and oxygen dissociation in preheated atmospheric pressure air subjected to nanosecond repetitively pulsed discharges. Figure~\ref{Rustholtz_domain} illustrates the two‑dimensional axisymmetric domain with a pin‑to‑pin electrode configuration. The electrodes are separated by a 4 mm gap, and each electrode has a radius of 1 mm, with the left electrode serving as the cathode and the right electrode as the anode. The time‑dependent voltage waveform $V(t)$, which has a peak voltage of 5.7 kV, is applied at the anode following the conditions reported by Rusterholtz~\textit{et al.}~\cite{rusterholtz2013ultrafast}. Ambient air enters the domain through the inlet at 2.6 m/s, and a zero‑gradient condition is applied at the outlet. For the transport
coefficients, the same equation, Eq.~\ref{transport_photo}, is used to obtain the electron mobility and diffusion coefficients. Following Hagelaar and Pitchford~\cite{hagelaar2005solving}, we approximate the electron energy transport coefficients as $\mu_\epsilon = \frac{5}{3} \mu_e$ and $D_\epsilon = \frac{5}{3} D_e$, which introduces a small error for nitrogen. Ion mobility is neglected due to the very short duration of the pulse discharge. All schemes are selected based on considerations of numerical stability: A first-order Euler scheme is employed for temporal discretization. The gradient terms are discretized using the Gauss linear scheme, while the divergence terms for the convection of the flow, species, electron density, and electron energy are discretized using the first-order upwind scheme. The viscous stress term is discretized using the Gauss linear scheme. Diffusion terms are evaluated using the Gauss linear corrected scheme, and linear interpolation is used for face-centered quantities. A corrected scheme is employed for the surface-normal gradients to improve the accuracy of the discretization on non-orthogonal meshes. Note that Poisson's equation for the electrostatic potential and the Helmholtz equations for photoionization are solved using the GAMG (Geometric-Algebraic Multi-Grid) solver, which is well suited for solving large sparse linear systems arising from elliptic equations. The computational mesh consists of 586{,}883 cells, and the minimum cell size between electrode is 3$\ \mu\text{m}$. Note that the same kinetic mechanism (32 species, 234 reactions) is employed in the result section. 

The temporal evolutions of \ce{N2(C)} and electron density reported by Rusterholtz~\textit{et al.}~\cite{rusterholtz2013ultrafast} were measured after the first pulse, during which the gas temperature between the electrodes reached approximately 1500 K, significantly higher than the ambient temperature of 1000 K. To consider the similar condition after the first pulse, we impose radial Gaussian profiles for the temperature and electron density across the gap~\cite{shao2025unified}: 
\begin{align}
\begin{split}
 & T(r,x) = 1000 + 500 \; \text{exp}\left(-r^2/\sigma_r^2\right), \\
 & n_e(r,x) = n_{bg} + n_{peak} \text{exp}\left(-r^2/\sigma_r^2\right),  
\end{split}
\end{align} 
where $\sigma_{r} = \num{225}\;\mu \mathrm{m}$, $n_{bg} = \num{1e15}/\mathrm{m^3}$, and $n_{peak}=\num{1.5e18}/\mathrm{m^3}$ are the values adopted in the present work. Note that $n_{peak}$ may vary depending on the electrode curvature, as Rusterholtz~\textit{et al.}~\cite{rusterholtz2013ultrafast} do not provide information about the electrode curvature. 

Figure~\ref{Rustholtz_streamer} showcases the temporal evolution of the electron density and electric field at three time instants: 0.2, 1.0, and 2.0 ns. At 0.2 ns, streamers initiate and propagate from the electrode tips, which results in a high-intensity electric field at the ionization front. Note that pre-existing plasma channels ($n_e \approx 10^{18}$ m$^{-3}$) from previous cycles influence the local electric field distribution, as a residual electric field is already established between the gaps. By 1.0 ns, the positive and negative streamers advance toward one another; the local electric field at the streamer heads also progresses. By 2.0 ns, the gap is fully bridged by a high-density plasma filament, forming a conductive channel that smooths the electric field. In Figure~\ref{Rustholtz_validation}, the temporal evolution of the maximum gas temperature, electron number density, and \ce{N2(C)} number density is compared with experimental data from Rusterholtz~\textit{et al.}~\cite{rusterholtz2013ultrafast}. Since our solver is capable of capturing the glow discharge phase, we employ the same pulse profiles up to 8.5~ns and set the applied voltage to zero thereafter. As shown in the temperature profile, both the numerical and experimental results exhibit the ultra-fast heating due to the rapid quenching of electronically excited nitrogen states.  For the electron number density, the numerical model predicts an immediate plateau at approximately $2 \times 10^{15} \text{ cm}^{-3}$. The slight overestimation at early times ($t < 5$ ns) compared to the experimental data is due to the initialized state of the preheated gas, which facilitates rapid ionization. The \ce{N_2(C)} density strongly depends on the pulse duration, reaching a peak at the end of the voltage application ($8.5$ ns) before rapidly decaying. The decay rate of \ce{N_2(C)} is consistent with the kinetic mechanisms governing energy transfer from electronically excited states, which drives the ultrafast gas heating observed in the temperature profiles.

\bibliographystyle{unsrt} 
\bibliography{cas-refs}

@article{markosyan2013high,
  title={High-order fluid model for streamer discharges: II. Numerical solution and investigation of planar fronts},
  author={Markosyan, AH and Dujko, S and Ebert, U},
  journal={Journal of Physics D: Applied Physics},
  volume={46},
  number={47},
  pages={475203},
  year={2013},
  publisher={IOP Publishing}
}

@article{taneja20241d,
  title={1D modeling of plasma streamers at ammonia-air flame conditions},
  author={Taneja, Taaresh Sanjeev and Sitaraman, Hariswaran and Yang, Suo},
  journal={Journal of Physics D: Applied Physics},
  volume={58},
  number={1},
  pages={015204},
  year={2024},
  publisher={IOP Publishing}
}

@article{zhelezniak1982photoionization,
  title={Photoionization of nitrogen and oxygen mixtures by radiation from a gas discharge},
  author={Zhelezniak, MB and Mnatsakanian, A Kh and Sizykh, Sergei Vasil’evich},
  journal={High Temperature Science},
  volume={20},
  number={3},
  pages={357--362},
  year={1982}
}

@article{luque2007photoionization,
  title={Photoionization in negative streamers: Fast computations and two propagation modes},
  author={Luque, Alejandro and Ebert, Ute and Montijn, Carolynne and Hundsdorfer, Willem},
  journal={Applied physics letters},
  volume={90},
  number={8},
  year={2007},
  publisher={AIP Publishing}
}

@article{bourdon2007efficient,
  title={Efficient models for photoionization produced by non-thermal gas discharges in air based on radiative transfer and the Helmholtz equations},
  author={Bourdon, Anne and Pasko, VP and Liu, Ningyu Y and C{\'e}lestin, S{\'e}bastien and S{\'e}gur, Pierre and Marode, Emmanuel},
  journal={Plasma Sources Science and Technology},
  volume={16},
  number={3},
  pages={656},
  year={2007},
  publisher={IOP Publishing}
}

@article{bagheri2018comparison,
  title={Comparison of six simulation codes for positive streamers in air},
  author={Bagheri, Behnaz and Teunissen, Jannis and Ebert, Ute and Becker, Markus M and Chen, She and Ducasse, Olivier and Eichwald, Olivier and Loffhagen, Detlef and Luque, Alejandro and Mihailova, Diana and others},
  journal={Plasma Sources Science and Technology},
  volume={27},
  number={9},
  pages={095002},
  year={2018},
  publisher={IOP Publishing}
}

@article{teunissen2017simulating,
  title={Simulating streamer discharges in 3D with the parallel adaptive Afivo framework},
  author={Teunissen, Jannis and Ebert, Ute},
  journal={Journal of Physics D: Applied Physics},
  volume={50},
  number={47},
  pages={474001},
  year={2017},
  publisher={IOP Publishing}
}

@article{rusterholtz2013ultrafast,
  title={Ultrafast heating and oxygen dissociation in atmospheric pressure air by nanosecond repetitively pulsed discharges},
  author={Rusterholtz, Diane L and Lacoste, Deanna A and Stancu, Gabi D and Pai, David Z and Laux, Christophe O},
  journal={Journal of Physics D: Applied Physics},
  volume={46},
  number={46},
  pages={464010},
  year={2013},
  publisher={IOP Publishing}
}

@article{jasak2009openfoam,
  title={OpenFOAM: Open source CFD in research and industry},
  author={Jasak, Hrvoje},
  journal={International journal of naval architecture and ocean engineering},
  volume={1},
  number={2},
  pages={89--94},
  year={2009},
  publisher={Elsevier}
}

@misc{reactingDNS_OpenFOAM,
  author       = {Zhensu She and collaborators},
  title        = {reactingDNS\_OpenFOAM: A DNS solver for reacting flows based on OpenFOAM},
  year         = {2020},
  howpublished = {\url{https://github.com/ZSHtju/reactingDNS_OpenFOAM}},
  note         = {Accessed: 2025-05-18}
}

@inproceedings{alkandry2013comparison,
  title={Comparison of models for mixture transport properties for numerical simulations of ablative heat-shields},
  author={Alkandry, Hicham and Boyd, Iain and Martin, Alexandre},
  booktitle={51st AIAA Aerospace Sciences Meeting including the New Horizons Forum and Aerospace Exposition},
  pages={303},
  year={2013}
}

@inproceedings{hash2007fire,
  title={FIRE II calculations for hypersonic nonequilibrium aerothermodynamics code verification: DPLR, LAURA, and US3D},
  author={Hash, David and Olejniczak, Joseph and Wright, Michael and Prabhu, Dinesh and Pulsonetti, Maria and Hollis, Brian and Gnoffo, Peter and Barnhardt, Michael and Nompelis, Ioannis and Candler, Graham},
  booktitle={45th AIAA Aerospace Sciences Meeting and Exhibit},
  pages={605},
  year={2007}
}

@article{franklin2003basic,
  title={Basic relationships in electronegative plasmas with a multiplicity of negative and positive ion species},
  author={Franklin, RN},
  journal={Journal of Physics D: Applied Physics},
  volume={36},
  number={7},
  pages={823--827},
  year={2003}
}

@article{prager2007modeling,
  title={Modeling ion chemistry and charged species diffusion in lean methane--oxygen flames},
  author={Prager, J and Riedel, U and Warnatz, J},
  journal={Proceedings of the Combustion Institute},
  volume={31},
  number={1},
  pages={1129--1137},
  year={2007},
  publisher={Elsevier}
}

@article{zhong2020structure,
  title={Structure and propagation of n-heptane/air premixed flame in low temperature ignition regime},
  author={Zhong, Shenghui and Zhang, Fan and Jangi, Mehdi and Bai, Xue-Song and Yao, Mingfa and Peng, Zhijun},
  journal={Applied Energy},
  volume={275},
  pages={115320},
  year={2020},
  publisher={Elsevier}
}

@article{zhu2017nanosecond,
  title={Nanosecond surface dielectric barrier discharge in atmospheric pressure air: I. measurements and 2D modeling of morphology, propagation and hydrodynamic perturbations},
  author={Zhu, Yifei and Shcherbanev, Sergey and Baron, Brian and Starikovskaia, Svetlana},
  journal={Plasma sources science and technology},
  volume={26},
  number={12},
  pages={125004},
  year={2017},
  publisher={IOP Publishing}
}

@article{sun2023detonationfoam,
  title={detonationFoam: An open-source solver for simulation of gaseous detonation based on OpenFOAM},
  author={Sun, Jie and Wang, Yiqing and Tian, Baolin and Chen, Zheng},
  journal={Computer Physics Communications},
  volume={292},
  pages={108859},
  year={2023},
  publisher={Elsevier}
}

@article{shao2025unified,
  title={A unified fluid model for nonthermal plasmas and reacting flows},
  author={Shao, Xiao and Lacoste, Deanna A and Im, Hong G},
  journal={Plasma Sources Science and Technology},
  volume={34},
  number={8},
  pages={085016},
  year={2025},
  publisher={IOP Publishing}
}

@article{zhong2023understanding,
  title={Understanding non-equilibrium N2O/NOx chemistry in plasma-assisted low-temperature NH3 oxidation},
  author={Zhong, Hongtao and Mao, Xingqian and Liu, Ning and Wang, Ziyu and Ombrello, Timothy and Ju, Yiguang},
  journal={Combustion and Flame},
  volume={256},
  pages={112948},
  year={2023},
  publisher={Elsevier}
}

@article{kortshagen2016nonthermal,
  title={Nonthermal plasma synthesis of nanocrystals: fundamental principles, materials, and applications},
  author={Kortshagen, Uwe R and Sankaran, R Mohan and Pereira, Rui N and Girshick, Steven L and Wu, Jeslin J and Aydil, Eray S},
  journal={Chemical reviews},
  volume={116},
  number={18},
  pages={11061--11127},
  year={2016},
  publisher={ACS Publications}
}

@article{ju2016plasma,
  title={Plasma assisted low temperature combustion},
  author={Ju, Yiguang and Lefkowitz, Joseph K and Reuter, Christopher B and Won, Sang Hee and Yang, Xueliang and Yang, Suo and Sun, Wenting and Jiang, Zonglin and Chen, Qi},
  journal={Plasma Chemistry and Plasma Processing},
  volume={36},
  number={1},
  pages={85--105},
  year={2016},
  publisher={Springer}
}

@article{ju2015plasma,
  title={Plasma assisted combustion: Progress, challenges, and opportunities},
  author={Ju, Yiguang and Sun, Wenting},
  journal={Combustion and Flame},
  volume={162},
  number={3},
  pages={529--532},
  year={2015},
  publisher={Elsevier}
}

@article{popov2016kinetics,
  title={Kinetics of plasma-assisted combustion: effect of non-equilibrium excitation on the ignition and oxidation of combustible mixtures},
  author={Popov, NA},
  journal={Plasma Sources Science and Technology},
  volume={25},
  number={4},
  pages={043002},
  year={2016},
  publisher={IOP Publishing}
}

@article{zhang2017non,
  title={Non-thermal plasma technology for organic contaminated soil remediation: A review},
  author={Zhang, Hao and Ma, Danyan and Qiu, Rongliang and Tang, Yetao and Du, Changming},
  journal={Chemical Engineering Journal},
  volume={313},
  pages={157--170},
  year={2017},
  publisher={Elsevier}
}

@article{schiavon2017non,
  title={Non-thermal plasma as an innovative option for the abatement of volatile organic compounds: a review},
  author={Schiavon, Marco and Torretta, Vincenzo and Casazza, Andrea and Ragazzi, Marco},
  journal={Water, Air, \& Soil Pollution},
  volume={228},
  number={10},
  pages={388},
  year={2017},
  publisher={Springer}
}

@article{lin2015microplasma,
  title={Microplasma: a new generation of technology for functional nanomaterial synthesis},
  author={Lin, Liangliang and Wang, Qi},
  journal={Plasma Chemistry and Plasma Processing},
  volume={35},
  number={6},
  pages={925--962},
  year={2015},
  publisher={Springer}
}

@article{colombo2008thermodynamic,
  title={Thermodynamic and transport properties in non-equilibrium argon, oxygen and nitrogen thermal plasmas},
  author={Colombo, Vittorio and Ghedini, Emanuele and Sanibondi, Paolo},
  journal={Progress in Nuclear Energy},
  volume={50},
  number={8},
  pages={921--933},
  year={2008},
  publisher={Elsevier}
}

@article{palma2020review,
  title={A review about the recent advances in selected nonthermal plasma assisted solid--gas phase chemical processes},
  author={Palma, Vincenzo and Cortese, Marta and Renda, Simona and Ruocco, Concetta and Martino, Marco and Meloni, Eugenio},
  journal={Nanomaterials},
  volume={10},
  number={8},
  pages={1596},
  year={2020},
  publisher={MDPI}
}

@article{chen2017characteristics,
  title={Characteristics and applications of plasma assisted chemical processes and reactors},
  author={Chen, Xin and Cheng, Yan and Li, Tianyang and Cheng, Yi},
  journal={Current Opinion in Chemical Engineering},
  volume={17},
  pages={68--77},
  year={2017},
  publisher={Elsevier}
}

@book{ju2025plasma,
  title={Plasma assisted combustion and chemical processing},
  author={Ju, Yiguang and Starikovskiy, Andrey},
  year={2025},
  publisher={CRC press}
}

@article{nijdam2020physics,
  title={The physics of streamer discharge phenomena},
  author={Nijdam, Sander and Teunissen, Jannis and Ebert, Ute},
  journal={Plasma Sources Science and Technology},
  volume={29},
  number={10},
  pages={103001},
  year={2020},
  publisher={IOP Publishing}
}

@article{bruggeman2017foundations,
  title={Foundations of atmospheric pressure non-equilibrium plasmas},
  author={Bruggeman, Peter J and Iza, Felipe and Brandenburg, Ronny},
  journal={Plasma Sources Science and Technology},
  volume={26},
  number={12},
  pages={123002},
  year={2017},
  publisher={IOP Publishing}
}

@article{huiskamp2020nanosecond,
  title={Nanosecond pulsed streamer discharges Part I: Generation, source-plasma interaction and energy-efficiency optimization},
  author={Huiskamp, T},
  journal={Plasma Sources Science and Technology},
  volume={29},
  number={2},
  pages={023002},
  year={2020},
  publisher={IOP Publishing}
}

@article{zhao2021elongation,
  title={Elongation and branching of stem channels produced by positive streamers in long air gaps},
  author={Zhao, Xiangen and Becerra, Marley and Yang, Yongchao and He, Junjia},
  journal={Scientific reports},
  volume={11},
  number={1},
  pages={4120},
  year={2021},
  publisher={Nature Publishing Group UK London}
}

@article{jacobsen2008plasma,
  title={Plasma-assisted ignition in scramjets},
  author={Jacobsen, Lance S and Carter, Campbell D and Jackson, Thomas A and Williams, Skip and Barnett, Jack and Bivolaru, Daniel and Kuo, Spencer and Tam, Chung-Jen and Baurle, Robert A},
  journal={Journal of Propulsion and Power},
  volume={24},
  number={4},
  pages={641--654},
  year={2008}
}

@article{do2010plasma,
  title={Plasma assisted cavity flame ignition in supersonic flows},
  author={Do, Hyungrok and Cappelli, Mark A and Mungal, M Godfrey},
  journal={Combustion and Flame},
  volume={157},
  number={9},
  pages={1783--1794},
  year={2010},
  publisher={Elsevier}
}

@article{ombrello2024scramjet,
  title={Scramjet cavity ignition using nanosecond-pulsed high-frequency discharges},
  author={Ombrello, Timothy M and Hammack, Stephen D and Carter, Campbell D and Busby, Kenneth and Lefkowitz, Joseph K},
  journal={Combustion and Flame},
  volume={262},
  pages={113335},
  year={2024},
  publisher={Elsevier}
}

@inproceedings{shen2025nanosecond,
  title={Nanosecond Pulsed Plasma Discharges in High-Speed Methane-Air Flow: Effects of Velocity on Plasma Morphology and Ignition Development},
  author={Shen, Si and Senior-Tybora, Weronika and Lefkowitz, Joseph},
  booktitle={AIAA SCITECH 2025 Forum},
  pages={0168},
  year={2025}
}

@article{hagelaar2016brief,
  title={Brief documentation of BOLSIG+ version 03/2016},
  author={Hagelaar, GJM},
  journal={Laboratoire Plasma et Conversion dEnergie (LAPLACE), Universit Paul Sabatier},
  volume={118},
  year={2016}
}

@article{hagelaar2005solving,
  title={Solving the Boltzmann equation to obtain electron transport coefficients and rate coefficients for fluid models},
  author={Hagelaar, GJM and Pitchford, L C},
  journal={Plasma sources science and technology},
  volume={14},
  number={4},
  pages={722--733},
  year={2005}
}

@article{eichwald2012effects,
  title={Effects of numerical and physical anisotropic diffusion on branching phenomena of negative-streamer dynamics},
  author={Eichwald, Olivier and Bensaad, Hanadi and Ducasse, Olivier and Yousfi, Mohammed},
  journal={Journal of Physics D: Applied Physics},
  volume={45},
  number={38},
  pages={385203},
  year={2012},
  publisher={IOP Publishing}
}

@article{bouwman20223d,
  title={3D particle simulations of positive air--methane streamers for combustion},
  author={Bouwman, Dennis and Teunissen, Jannis and Ebert, Ute},
  journal={Plasma Sources Science and Technology},
  volume={31},
  number={4},
  pages={045023},
  year={2022},
  publisher={IOP Publishing}
}

@article{pancheshnyi2008zdplaskin,
  title={ZDPlasKin: a new tool for plasmachemical simulations},
  author={Pancheshnyi, Sergey and Eismann, Benjamin and Hagelaar, Gerjan and Pitchford, Leanne},
  journal={Bulletin of the American Physical Society},
  volume={53},
  pages={1--14},
  year={2008},
  publisher={American Physical Society}
}

\end{document}